\documentclass[12pt]{article}
\usepackage{graphicx,amssymb,bbold}
\usepackage{amsmath,amsfonts,epsfig}
\usepackage{subfig}
\newcommand{\be}{\begin{equation}}
\newcommand{\ee}{\end{equation}}
\newcommand{\bea}{\begin{eqnarray}}
\newcommand{\eea}{\end{eqnarray}}
\newcommand{\beas}{\begin{eqnarray*}}
\newcommand{\eeas}{\end{eqnarray*}}

\usepackage{color}

\begin{document}
\begin{titlepage}

\vspace*{-24mm}
\rightline{YITP-15-17}
\vspace{4mm}

\begin{center}

{\bf {\Large  
Quantum Black Hole Formation}} \\[3pt] 
\vspace{1mm}
{\bf {\Large in the BFSS Matrix Model}} 

\vspace{8mm}

\renewcommand\thefootnote{\mbox{$\fnsymbol{footnote}$}}
Sinya {\sc Aoki}${}^{a}$\footnote{saoki@yukawa.kyoto-u.ac.jp},  
Masanori {\sc Hanada}${}^{abc}$\footnote{hanada@yukawa.kyoto-u.ac.jp} 
and
Norihiro {\sc Iizuka}${}^{d}$\footnote{iizuka@phys.sci.osaka-u.ac.jp}

\vspace{4mm}

${}^a${\small \sl Yukawa Institute for Theoretical Physics} \\
{\small \sl Kyoto University, Kyoto 606-8502, JAPAN}

${}^b${\small \sl The Hakubi Center for Advanced Research} \\ 
{\small \sl Kyoto University, 
Kyoto 606-8501, JAPAN}

${}^c${\small \sl Stanford Institute for Theoretical Physics} \\
{\small \sl Stanford University, Stanford, CA 94305, USA}

${}^d${\small \sl Department of Physics, Osaka University} \\ 
{\small \sl Toyonaka, Osaka 560-0043, JAPAN}

\end{center}

\vspace{4mm}

\noindent
We study the various head-on collisions of two bunches of D0-branes 
and their real-time evolution in the BFSS matrix model in classical limit.  
For a various matrix size $N$ respecting the 't Hooft scaling, 
we find quantitative evidence for the formation of  a single bound state of D0-branes at late time, 
which is matrix model thermalization and dual to the formation of a  larger black hole.

\end{titlepage}

\setcounter{footnote}{0}
\renewcommand\thefootnote{\mbox{\arabic{footnote}}}

\section{Introduction\label{sect:intro}}
Black holes have always been at the centre of theoretical interests in quantum gravity and string theory. 
Through the non-perturbative formulation of string theory and quantum gravity via gauge/string correspondence \cite{Maldacena:1997re},   
significant progresses have been made in our understanding of black hole microstates, especially 
its entropy counting. 
However, we still need to understand better various aspects of  
quantum gravitational system, and one of the important aspects 
is understanding their real-time evolution. 

One model, which is more tractable to analyse the real-time dynamics compared with many other theoretical models in the gauge/string correspondence, is a BFSS (Banks-Fischler-Shenker-Susskind) 
matrix model \cite{Banks:1996vh}. This model is a 0 + 1 dimensional large $N$ matrix quantum mechanics, and  
conjectured \cite{Itzhaki:1998dd, Polchinski:1999br} to be dual to 
either ten-dimensional string or eleven dimensional 
M theory depending on the various parameter limit.  
Since BFSS matrix model is dual to the higher dimensional spacetime, it would be very interesting to ask how the higher dimensional black hole formation can be understood from the dual matrix model. 
In this set-up, we consider the various head-on collisions of two bunches of D0-branes and their  
real-time evolution, which is expected to become a {\it single} bound state of D0-branes at late time. 
This is a matrix model thermalization and dual to the formation of one large black hole  
in either ten- or eleven dimension. Our main interest in this paper is, by solving the 
matrix model classical equation of motion, 
to study quantitatively how such a bound state is formed in the real-time evolution of the large $N$ matrix model.

In order to study the complicated time-evolution of large $N$ matrix model, 
we make several simplifications; 
First, we consider only the classical limit in this paper, namely we solve the classical equation of motion numerically and find its time-dependent solution. At first thought, this simplification might 
sound very boring limit. 
However, there are several reasons why classical limit of late time evolution is interesting  and capture aspects of non-perturbative physics.  
Even though classical limit is justified in the weak coupling limit, 
since what we conduct is not the perturbative analysis near the trivial vacuum but rather the analysis 
seeking the time-dependent soliton 
configuration, this can capture non-perturbative physics.   
Remember that thermalization is a very complicated but also universal phenomenon 
which one cannot see in various solvable models, 
nor in perturbation analysis from the trivial vacuum \cite{Festuccia:2006sa}. 
It has a property that all memories of the initial data are lost in late time, and 
{this late time thermalization may occur} only in the large $N$ limit.  
Therefore even though we simply solve the classical equations of motion, we are looking some non-perturbative physics through the late time evolution.

Another simplification is,  
that we {neglect effects of  fermions}  for simplicity. Since supersymmetry is expected to be 
more crucial at low energy but not  in high energy deconfined phase, we expect that, {for late time thermalization, }
the analysis with/without supersymmetry does not change much. 
In our analysis, we have conduct $N$ up to $N=24$, and regard it as large enough to {perform an extrapolation to}  the large $N$ limit.

Before we close the introduction, we comment on several related references. 
Thermal equilibrium (i.e., static) properties of the BFSS matrix models for black holes are very well studied.  
Thermodynamic simulations of BFSS matrix model 
was initiated by Kabat et al \cite{Kabat:2001ve,Kabat:2000zv} by 
the mean-field method \cite{Kabat:1999hp},  
and recently by Monte Carlo method \cite{Anagnostopoulos:2007fw,Catterall:2008yz,Catterall:2009xn}. 
These results confirmed that the duality is valid 
not just at supergravity level 
but also at stringy level; at finite-coupling \cite{Hanada:2008ez} and finite-$N$ \cite{Hanada:2013}.   
For non-equilibrium properties, a probe limit thermalization was studied in \cite{Iizuka:2001cw,Iizuka:2008hg,Berenstein:2013tya}. 
For so-called BMN matrix model\footnote{BMN matrix model \cite{Berenstein:2002jq} is a model which has mass for the adjoint scalars. This model is supposed to be dual to the 11-dimensional pp-waves.}, 
Berenstein et al studied the thermalization numerically in the series of papers 
 \cite{Berenstein:2010bi, Asplund:2011qj, Asplund:2012tg} in detail. 
BFSS matrix model with initial conditions different from ours have also been studied in \cite{Asplund:2011qj,Asplund:2012tg}.   
Finally the black hole formulation at the correspondence point for the BFSS matrix model was analytically studied recently in \cite{Iizuka:2013yla,Iizuka:2013kha}.  
In this paper we consider the similar setting to these works, though the parameter range is different, which is possible since we solve the model numerically.

\section{The model and simulation method}
BFSS matrix model \cite{Banks:1996vh} is the maximally supersymmetric matrix quantum mechanics, 
which is the dimensional reduction of 4d ${\cal N}=4$ supersymmetric Yang-Mills theory to one dimension (time). 
In this paper, we concentrate on the classical dynamics, and set fermion to zero for simplicity. 
Henceforth, we consider only the bosonic part of the Lagrangian of BFSS matrix model and it is given by 
\begin{eqnarray}
\label{BFSSaction}
L
=
\frac{N}{2g_{YM}^2 N}{\rm Tr}\left(
\sum_{M=1}^9(D_tX_M)^2
+
\frac{1}{2}
\sum_{M \neq N}[X_M,X_N]^2
\right). 
\end{eqnarray}
Here $X_M$ $(M=1,2,\cdots,9)$ are $N\times N$ Hermitian matrices and $(D_tX_M)$ is the covariant derivative given by 
$(D_tX_M)=\partial_t X_M-[A_t,X_M]$, where $A_t$ is the $U(N)$ gauge field. 
This matrix model is a gravity-decoupled theory on the $N$ D0-branes \cite{Witten:1995im}.  
Large $N$ bound state of D0-branes with finite temperature $T$ with large 't Hooft coupling 
is dual to a black hole (black 0-brane) with Hawking temperature $T$ \cite{Itzhaki:1998dd}. Such a bound state is described by highly non-commutative matrices. On the other hand,  if matrices are close to block-diagonal, it describes multi-D0-branes and each of which is highly quantum\footnote{In order to have classical black hole picture we need both large 't Hooft coupling and large $N$. For single D0-brane, $N=1$ therefore highly quantum description.} black hole.

In the $A_t=0$ gauge, the classical dynamics is described by the equation of motion 
\begin{eqnarray}
\frac{d^2X_M}{dt^2}
-
\sum_{N}[X_N,[X_M,X_N]]
=
0
\label{eq:EOM}
\end{eqnarray}
with the Gauss's law constraint 
\begin{eqnarray}
\sum_{M}\left[X_M,\frac{dX_M}{dt}\right]
=
0. 
\label{eq:constraint}
\end{eqnarray}

In the rest of this paper we solve this classical equation of motion. From the action (\ref{BFSSaction}) it is clear that 't Hooft coupling $\lambda\equiv g_{YM}^2N$ appears only 
as an overall factor. Therefore the classical limit, {\it i.e.,} $\hbar \to 0$ limit, is equivalent to 
the effective coupling constant $\lambda_{eff} \equiv \lambda/U^3$ goes to zero limit in the 't Hooft scaling limit, where $N \to \infty$ and $U$ is the typical VEV scale in the matrix model associated with the temperature, or quantum fluctuations of the matrix degrees of freedom. Therefore, our analysis of solving classical equation of motion is justified in 
hight temperature/energy limit\footnote{From the scaling symmetry 
$t \to t/\alpha$, $X_M \to \alpha X_M$ of classical equation of motion (\ref{eq:EOM}), one might wonder if we can re-scale the total energy. However this is a symmetry only at the classical level and full quantum theory does not possess such a symmetry.}.

\subsection{Discretization}
In order to study the time evolution, we discretize the equation of motion \eqref{eq:EOM} while preserving 
the constraint \eqref{eq:constraint} exactly. 
For that purpose, we write the EOM as 
\begin{eqnarray}
V_M(t)
=
\frac{dX_M(t)}{dt}, 
\qquad
F_M(t)
=
\frac{dV_M(t)}{dt} 
\end{eqnarray}
and 
\begin{eqnarray}
F_M(t)
=
\sum_{j}[X_N(t),[X_M(t),X_N(t)]]. 
\end{eqnarray}
The constraint becomes 
\begin{eqnarray}
\sum_{M}[X_M(t),V_M(t)]=0. 
\label{eq:constraint_regularized}
\end{eqnarray}
Under the infinitesimal time evolution $t\to t+\delta t$, $X_M$ and $V_M$ change as 
\begin{eqnarray}
X_M(t+\delta t)
=
X_M(t)
+
V_M(t)\cdot\delta t
+
F_M(t)\cdot\frac{(\delta t)^2}{2}
+
O\left((\delta t)^3\right), 
\end{eqnarray}
\begin{eqnarray}
V_M(t+\delta t)
=
V_M(t)
+
\left(
F_M(t)
+
F_M(t+\delta t)\right)\cdot\frac{\delta t}{2}
+
O\left((\delta t)^3\right), 
\end{eqnarray}
as one can easily check by using the standard Taylor expansion. 
We terminate this expansion at the order of $(\delta t)^2$, 
\begin{eqnarray}
X_M(t+\delta t)
\equiv
X_M(t)
+
V_M(t)\cdot\delta t
+
F_M(t)\cdot\frac{(\delta t)^2}{2}, 
\label{eq:EOM_regularized_1}
\end{eqnarray}
\begin{eqnarray}
V_M(t+\delta t)
\equiv
V_M(t)
+
\left(
F_M(t)
+
F_M(t+\delta t)\right)\cdot\frac{\delta t}{2}. 
\label{eq:EOM_regularized_2}
\end{eqnarray}
Then, a simple but tedious calculation shows that the constraint \eqref{eq:constraint_regularized} is satisfied at $t+\delta t$ 
if it holds at $t$, without any additional higher-order terms: 
\begin{eqnarray}
\sum_{M}[X_M(t),V_M(t)]=0
\quad
\Longrightarrow
\quad
\sum_{M}[X_M(t+\delta t),V_M(t+\delta t)]=0. 
\end{eqnarray}
Therefore, we use \eqref{eq:EOM_regularized_1} and \eqref{eq:EOM_regularized_2}, 
and choose the initial configuration to satisfy \eqref{eq:constraint_regularized} at $t=0$, 
\begin{eqnarray}
\sum_{M}[X_M(0),V_M(0)]=0.
\end{eqnarray}

\subsection{Initial condition -- collision of two bunches of D0-branes -- }
We consider a collision of two bunches of D0-branes, which is analogous to two black zero-branes, or two black holes (the left of Fig.~\ref{BH_formation}). 
The initial condition is\footnote{It would be better to introduce the off-diagonal elements also in the diagonal blocks. 
For simplicity, we do not do that for the moment. 
(If we introduce them, then the width should be taken larger than that for the off-diagonal blocks, $\sigma$.)} 
\begin{eqnarray}
& &
X_1=
\left(
\begin{array}{cc}
-\frac{L}{2}\cdot\textbf{1}_{N/2} & \\
 & \frac{L}{2}\cdot\textbf{1}_{N/2}
\end{array}
\right), 
\nonumber\\
& &
V_1=
\left(
\begin{array}{cc}
\frac{V}{2}\cdot\textbf{1}_{N/2} & \\
 & -\frac{V}{2}\cdot\textbf{1}_{N/2}
\end{array}
\right), 
\nonumber\\
& &
V_2=\cdots =V_9=0, 
\end{eqnarray}
and 
\begin{eqnarray}
X_\mu^{ij} = (X_\mu^{ji})^\ast = \sigma (a_{\mu,ij}+ib_{\mu,ij})
\end{eqnarray}
for $\mu=2,3,\cdots,9$, $i=1,2,\cdots, N/2$ and $j=N/2+1,\cdots,N$, 
where $a_{\mu,ij}$ and $b_{\mu,ij}$ are Gaussian random numbers with the weights $e^{-a^2/2}$ and $e^{-b^2/2}$. 
Here $\sigma$ is the source for off-diagonal elements and therefore non-commutativity,  which is crucial ingredients for the bound state formation. If $\sigma = 0$, then commutator square potential vanishes and two bunches just pass by. 
For nonzero $\sigma,$ the quantitative details of the initial condition can depend on the choice of random numbers, especially when $N$ is small.  
Note that we consider only even values of $N$. 
This initial condition satisfies the Gauss-law constraint $\sum_{M=1}^9[X_M,V_M]=0$. 
We consider the 't Hooft large-$N$ limit, where $\lambda\equiv g_{YM}^2N$ is fixed, 
$L$ and $V$ are fixed, and then, the energy scales as $N^2$. When two bunches are separated, 
the off-diagonal element $\sigma$ has larger mass as $\sigma$ and so quantum mechanically 
path-integrated out perturbatively in $\lambda/\sigma^3$ and suppressed. 
This describes the quantum fluctuation of open strings stretched between two bunches.  
Because the potential energy at $t=0$ is approximately 
$\frac{N}{\lambda}Tr[X_M,X_N]^2 \sim  N\cdot N^2L^2\sigma^2 $,   
the natural scaling is $\sigma\sim 1/(\sqrt{N}L)$. We will see that the energy actually scales as $E\sim N^2$ then. 

\section{Formation of a large quantum black hole}

\subsection{Formation of a single bound state}\label{sec:bunch_formation}

\begin{figure}[htb]
\begin{center}
\scalebox{0.3}{
\includegraphics{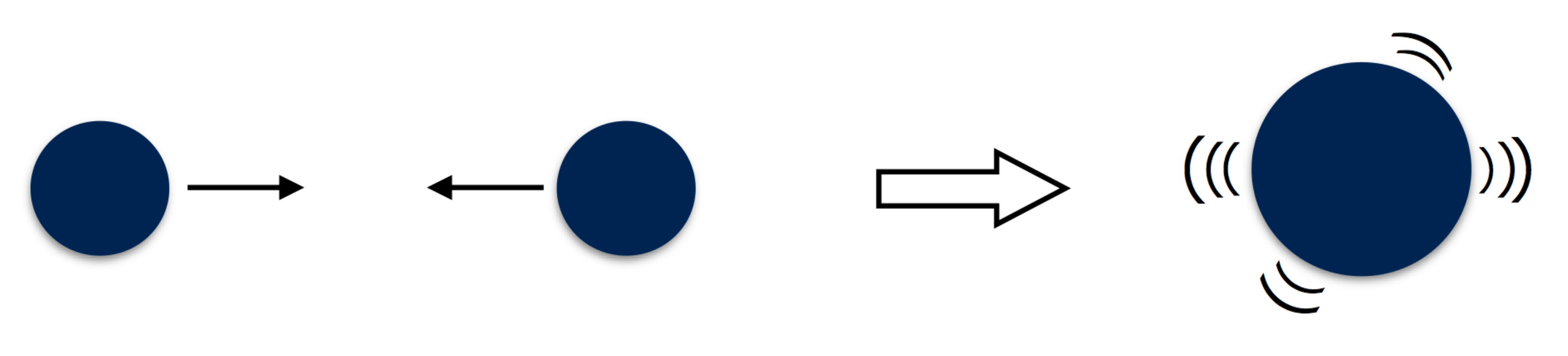}}
\end{center}
\caption{If a single bound state of eigenvalues is formed, it is analogous to a formation of 
one large black hole, 
and this is the dual to the thermalization process of the whole matrix system.}\label{BH_formation}
\end{figure}

Let us first ask `when a single bound state of eigenvalues is formed'.  
This is the process where two bunches of D0-branes, or equivalently two quantum black holes merge to form one large quantum black hole (Fig.~\ref{BH_formation}).  

\begin{figure}[htbt]
\centering  
\subfloat{\includegraphics[clip, width=4.0in]{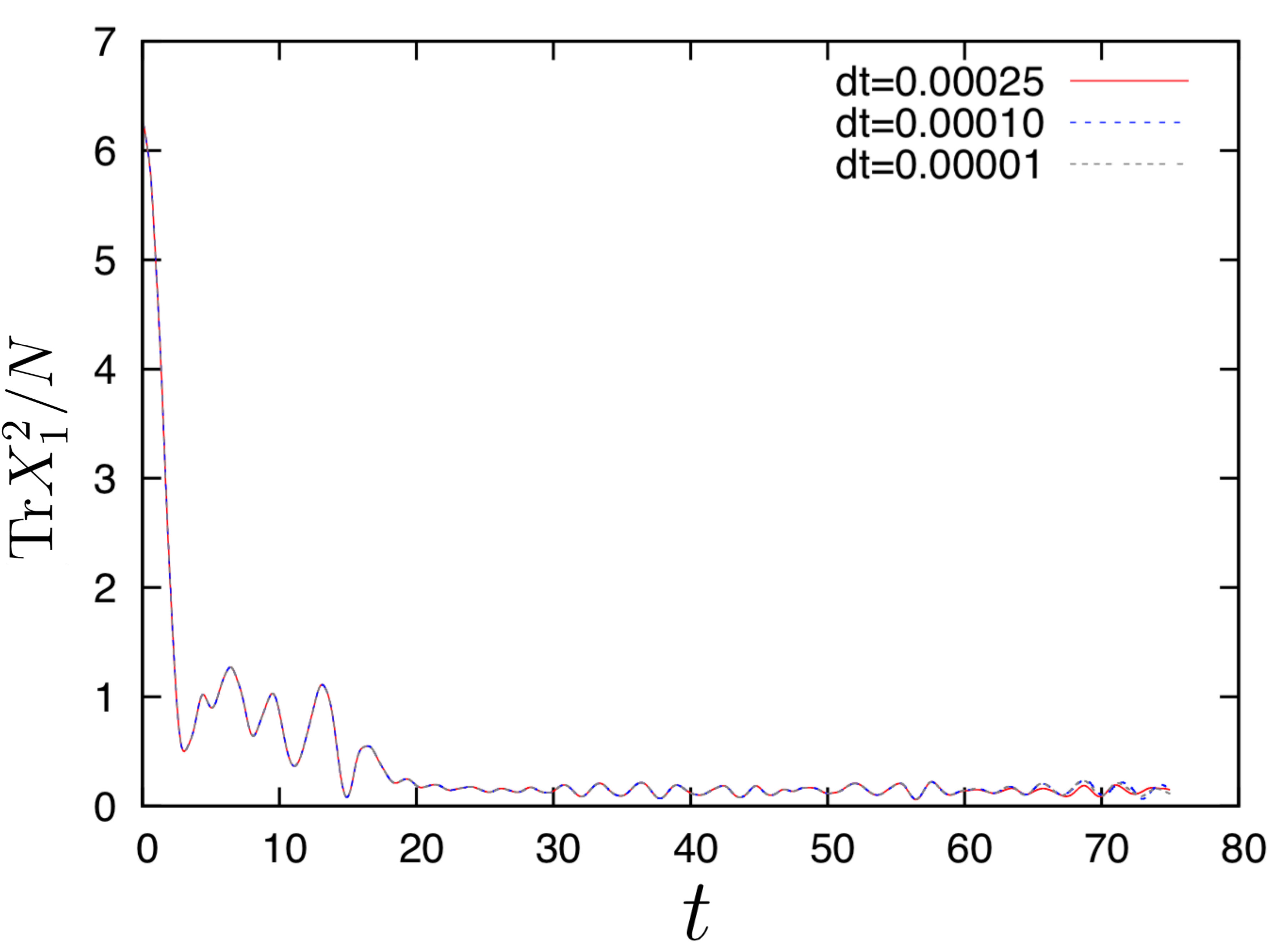}}
\\
 \subfloat{\includegraphics[clip, width=4.2in]{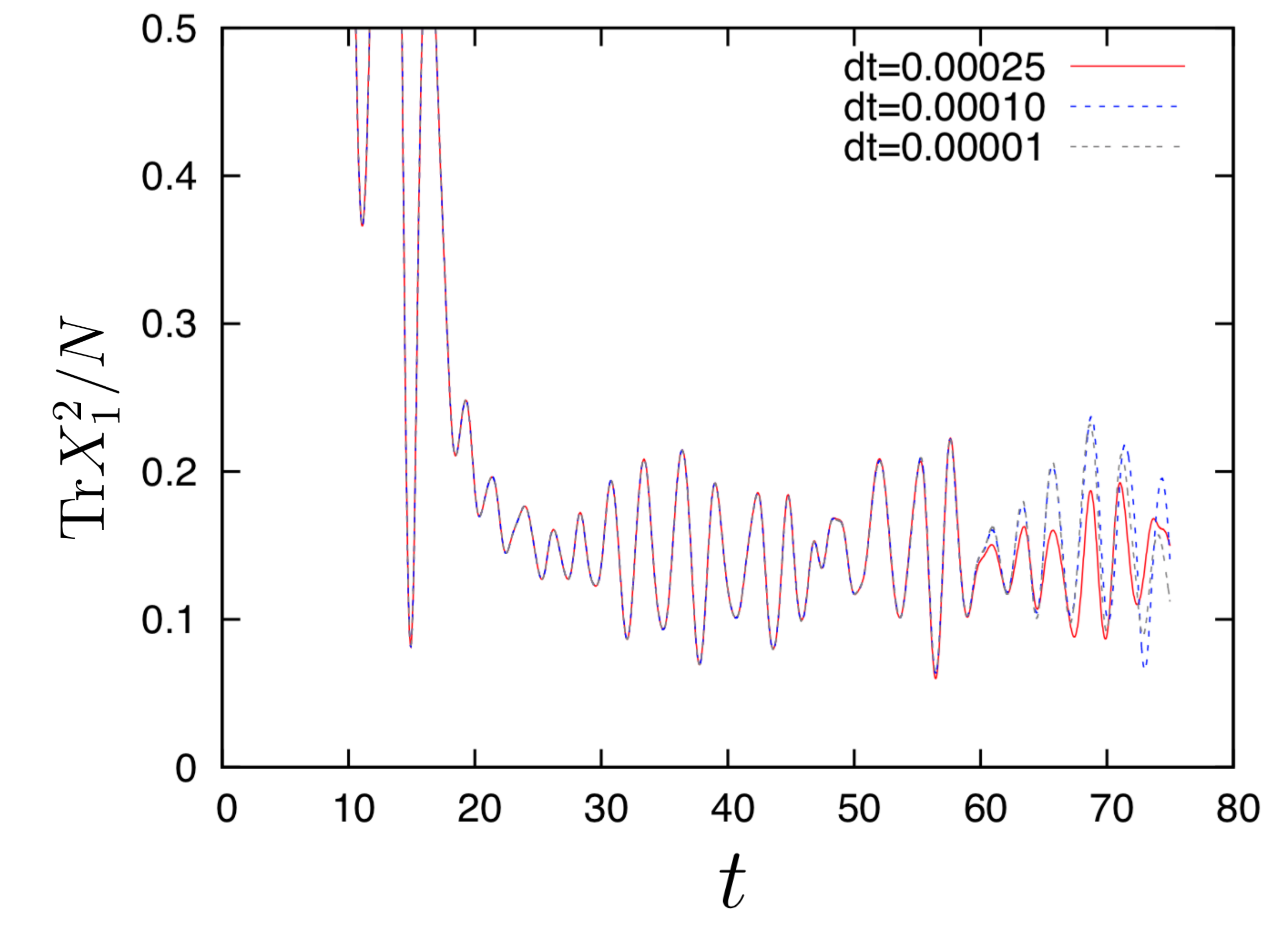}}
 \caption{${\rm Tr}X_1^2/N$ for $N=8$, for $L=5.0$, $V=0$ and $\sqrt{N}\sigma=0.12$; 
 $dt=0.00025, 0.00010$ and $0.00001$. 
 The horizontal axis is time $t$. We can see that $dt=0.00025$ gives correct answer up to $t\simeq 60$. 
 Beyond there, the error becomes large quickly, even for $dt=0.00010$. 
 The bottom panel is a zoom-up of the top one.}\label{trX1_N8}
\end{figure}

\begin{figure}[htb]
\begin{center}
\scalebox{0.27}{
\includegraphics{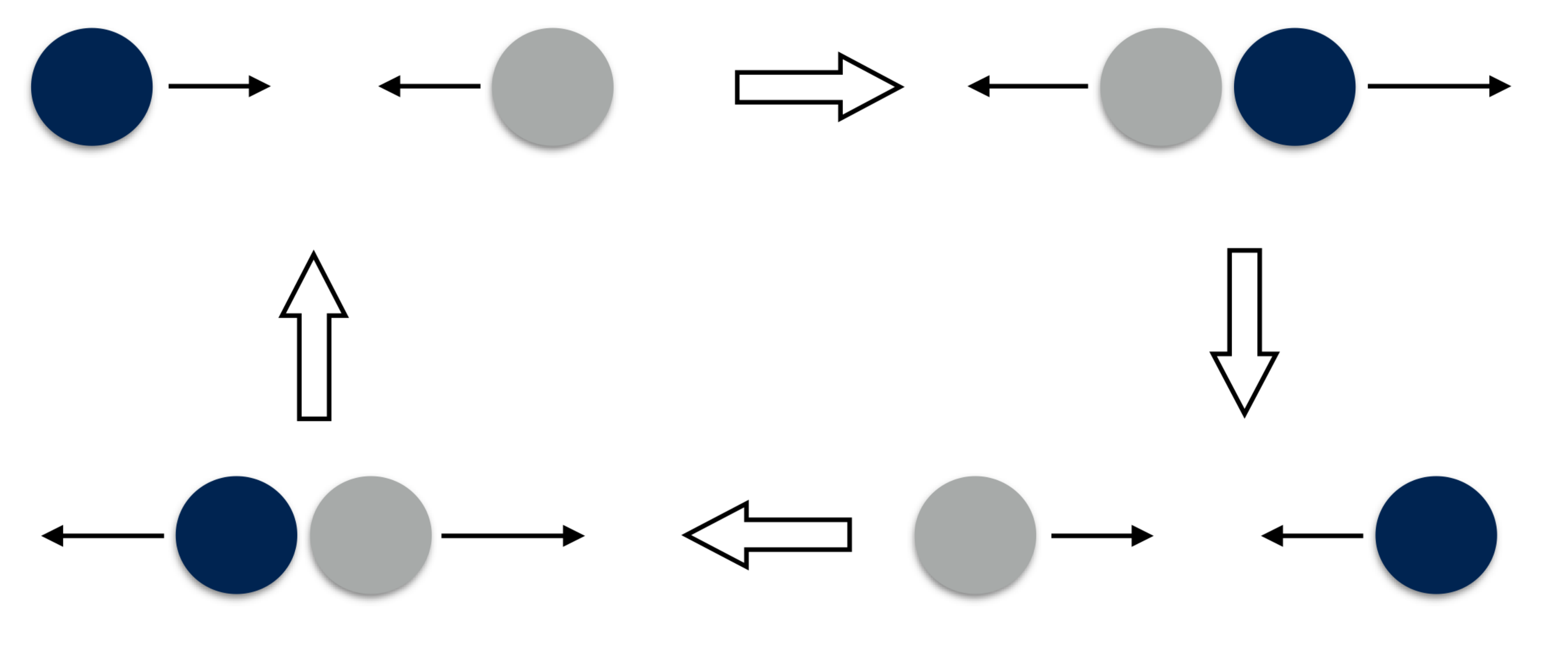}}
\end{center}
\caption{
A `bound state' would not necessarily be a single black hole; for example, it would be possible that two bunches oscillates at the late time. 
However such a configuration cannot be regarded as thermalized state. 
}\label{oscillation}
\end{figure}

The easiest way to see this process is to plot ${\rm Tr}X_M^2$, especially ${\rm Tr}X_1^2$, as a function of $t$. 
In Fig.~\ref{trX1_N8}, ${\rm Tr}X_1^2/N$ for $N=8$, for $L=5.0$, $V=0$ and $\sqrt{N}\sigma=0.12$, 
with several different values of time step $dt$, is shown. 
Here we set the 't Hooft scale $\lambda = 1$.  
At late time,  errors associated with discrete time steps become non-negligible. 
We can see that behaviors at $t\le 70$ can reliably be studied with $dt=0.00010$. 
${\rm Tr}X_1^2/N$  bounces a few times and then stays small, which suggests the formation of a single bound state.  
Note that, if instead two bunches oscillated as shown in Fig.~\ref{oscillation}, ${\rm Tr}X_1^2/N$ would oscillate without decaying.

\begin{figure}[htbt]
\centering  
\rotatebox{0}{
\subfloat{\includegraphics[clip, width=4.0in]{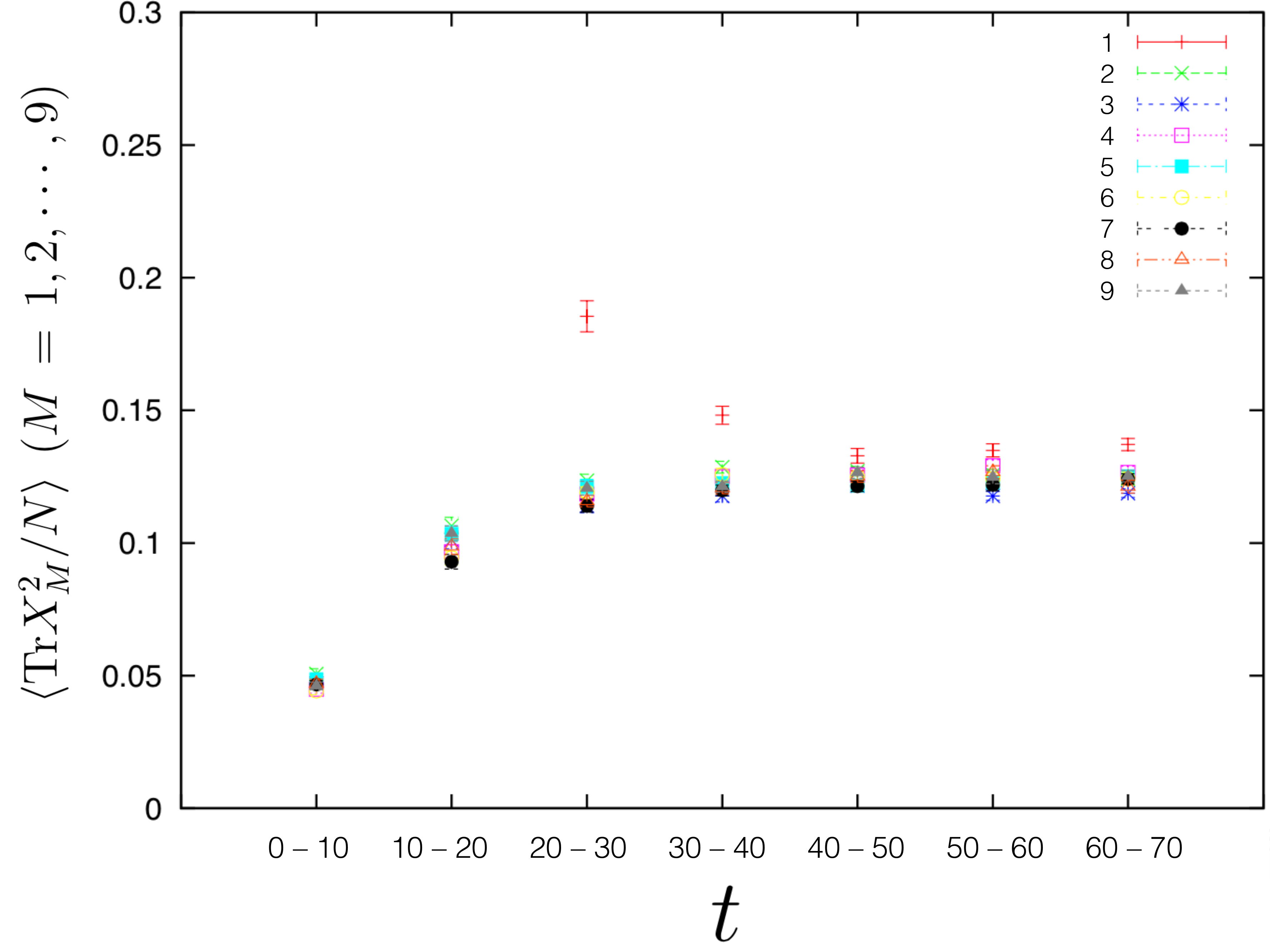}}}
\\
\rotatebox{0}{
 \subfloat{\includegraphics[clip, width=4.0in]{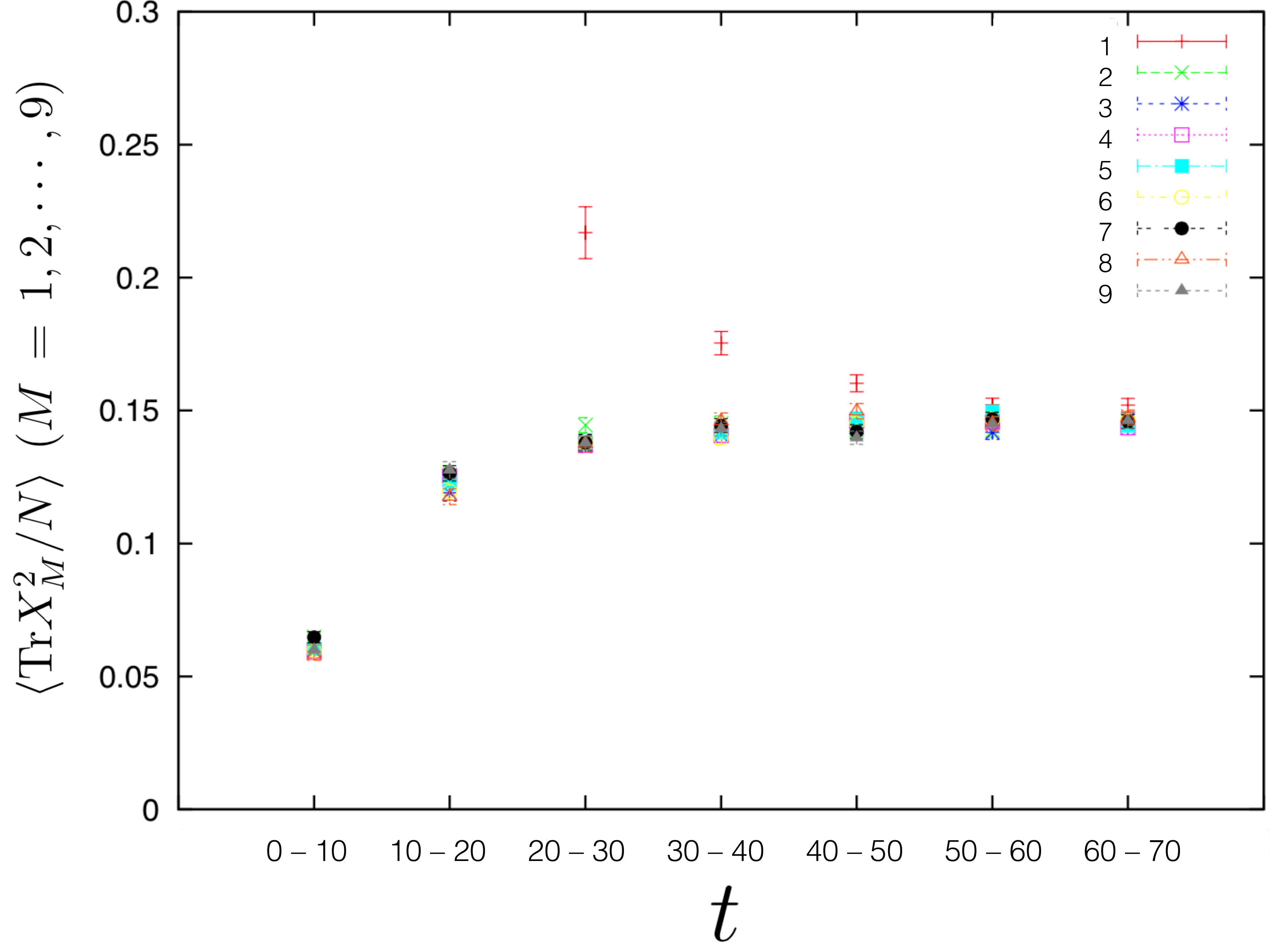}}}
\caption{$\langle {\rm Tr}X_M^2/N\rangle$ ($M=1,2,\cdots,9$), $N=8$, for $L=5.0$, $V=0$ and $dt=0.00010$,  
 $\sqrt{N}\sigma=0.12$ (top) and $\sqrt{N}\sigma=0.14$ (bottom). 
 The values along the horizontal axis are the time range used for taking the time average. 
 The error bar is purely statistical. We can see the restoration of the rotational symmetry at late time.  
 }\label{trX_N8_average}
\end{figure}

At late time, we expect that the system goes to the the ``typical'' configurations 
and that such typical states are rotationally symmetric after taking the average over time and/or different initial configurations. 
In Fig.~\ref{trX_N8_average}, the average $\langle TrX_M^2/N\rangle$ ($M=1,2,\cdots,9$), 
by using $0\le t\le 10$, $10\le t\le 20$, $\cdots$, $60\le t\le 70$, for 100 different initial configurations, are plotted. 
We can see that the average values converge to the same value at late time, suggesting the restoration of the rotational symmetry.

\subsubsection{Insensitivity to initial conditions for thermalization}

\begin{figure}[htbt]
\centering  
\subfloat{\includegraphics[clip, width=4.0in]{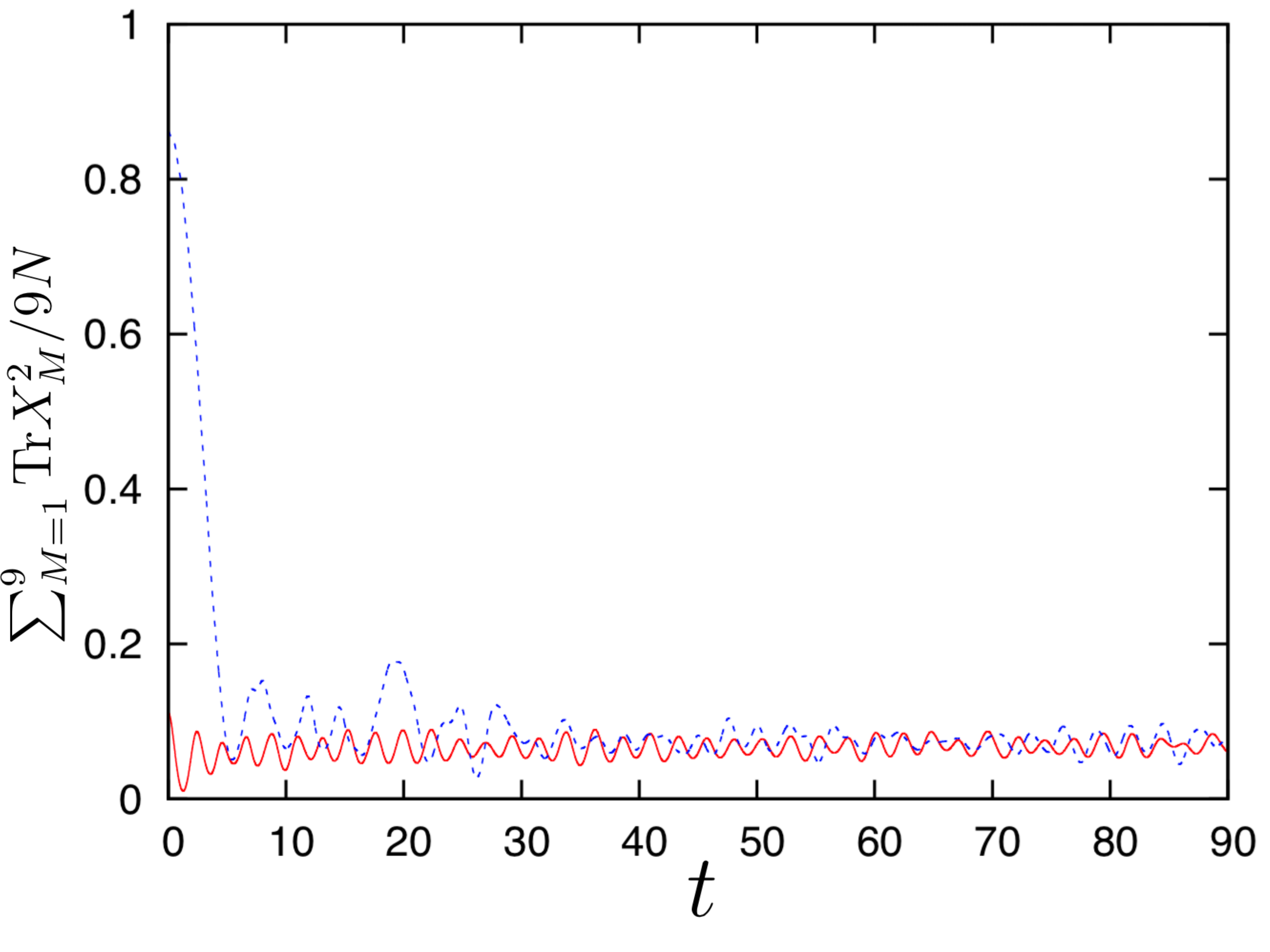}}
\\
 \subfloat{\includegraphics[clip, width=4.2in]{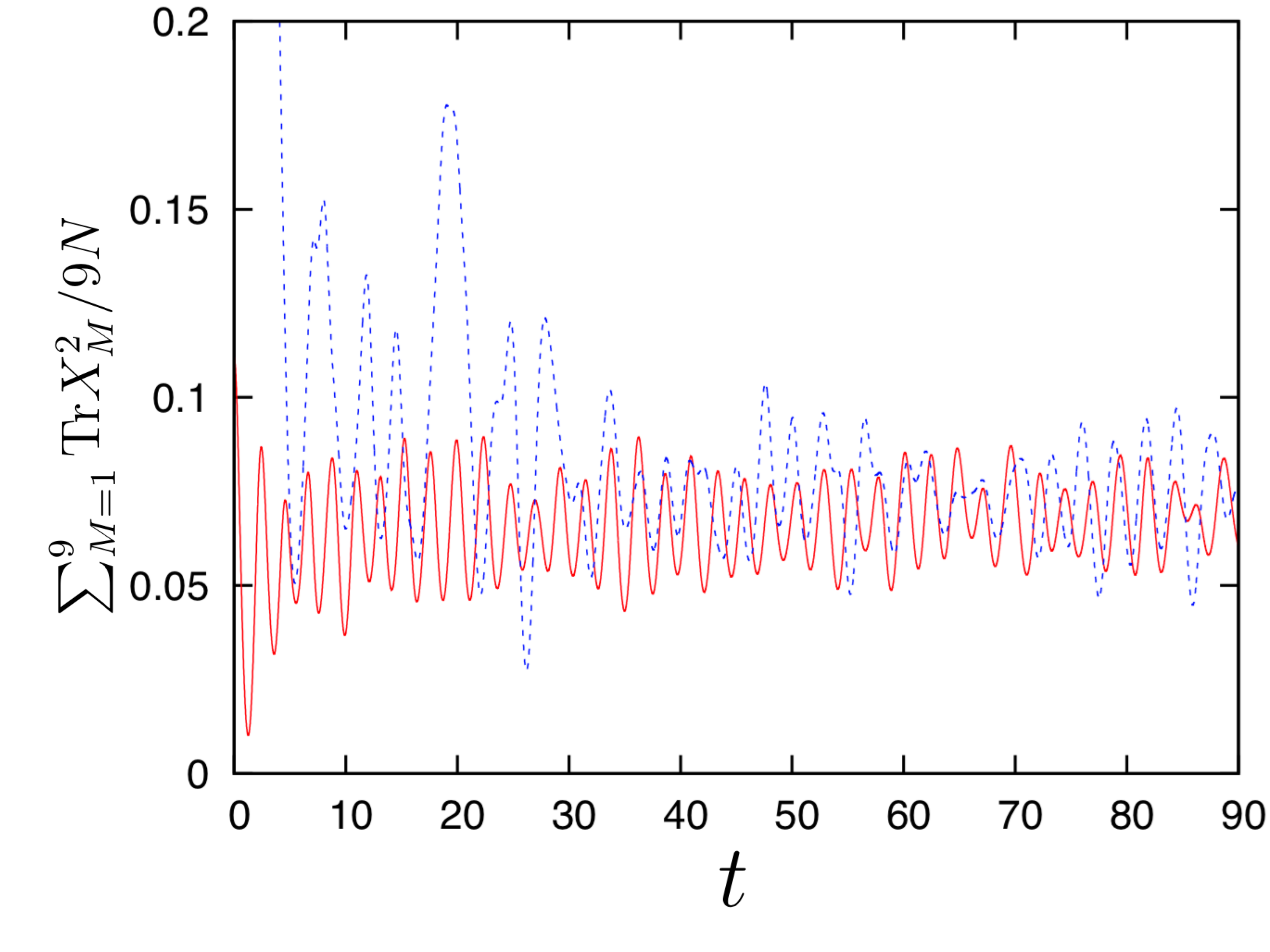}}
\caption{$\sum_{N=1}^9{\rm Tr}X_M^2/9N$, $N=4$, $E/N^2=0.5$, $dt=0.000001$. 
 Blue and red lines represent the initial conditions (i) and (ii) in the main text, respectively. 
 The bottom panel is the zoom up of the top one. 
 The late time behaviors are indistinguishable. 
  }\label{N4E05_TrX2}
\end{figure}

In order to add further evidence for thermalization, 
we studied two completely different initial conditions with the same value of the energy. 
In order to tune the total energy, we set $V_M=0$ and rescale $X_M$ to $\alpha X_M$. 
Then the energy scales as $E\to \alpha^4 E$. By using this, we can tune the energy to any value. 
Here we consider (i) $N=4$, $L=10.0$, $V=0$, $\sqrt{N}\sigma=0.12$, and (ii) $N=4$, all matrix components are generated Gaussian weight. Then we rescale $X_M$ so that $E/N^2=0.5$. 
Note that (i) is much more anisotropic initial condition than (ii). 
We employed $dt=0.00001$ and $dt=0.000001$ in this calculation, which give consistent results at $t<90$. 
From Fig.~\ref{N4E05_TrX2}, we can see that $\sum_{M=1}^9{\rm Tr}X_M^2/9N$  with two different initial conditions behave the same manner  in late time when 
the energies are the same, and it is hard to tell the initial condition unless we explicitly solve the equation of motion to go back to the past. These suggest typical thermliazation, {\it i.e.,}
whatever initial conditions it starts with, a system ends up with similar typical states.

\subsection{Large-$N$ limit and Thermalization}\label{sec:Virialand1/N}

In order to study the statistical nature of the BFSS matrix model, it is not essential to take the step size $dt$ very small, 
as long as the energy is conserved and we take enough late time evolution. 
In the following, we take $dt=0.0005$.

\subsubsection{Large-$N$ limit with fixed $L$, $V$ and $\sqrt{N}\sigma$}
Let us first consider a large-$N$ limit with fixed $L$, $V$ and $\sqrt{N}\sigma$. 
As a concrete example, we consider $L=5.0$, $V=0$ and $\sqrt{N}\sigma=0.12$. 
Similar results were obtained for other values too.

\begin{figure}[htbt]
\centering  
\subfloat{\includegraphics[clip, width=4.0in]{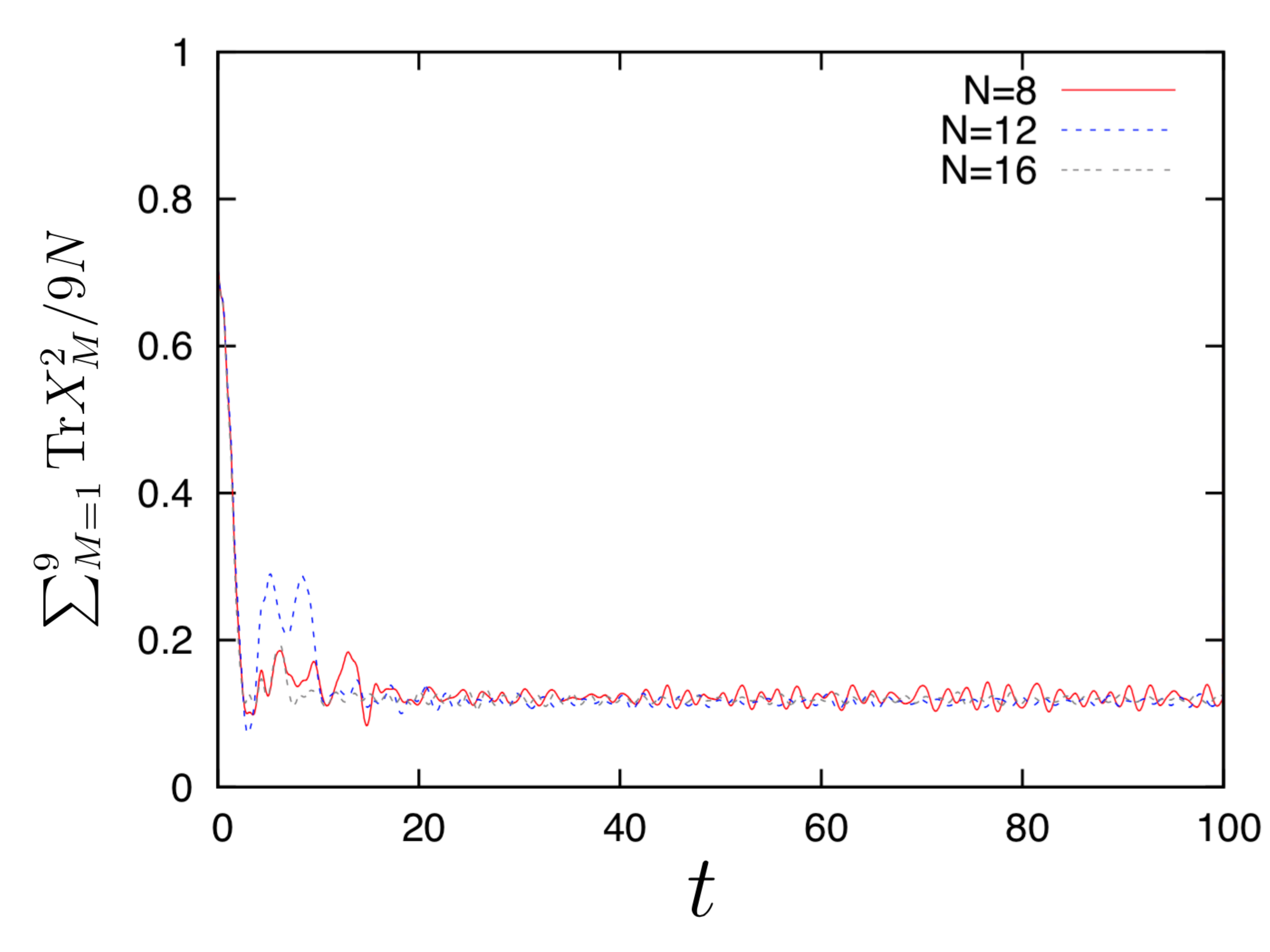}}
\\
 \subfloat{\includegraphics[clip, width=4.0in]{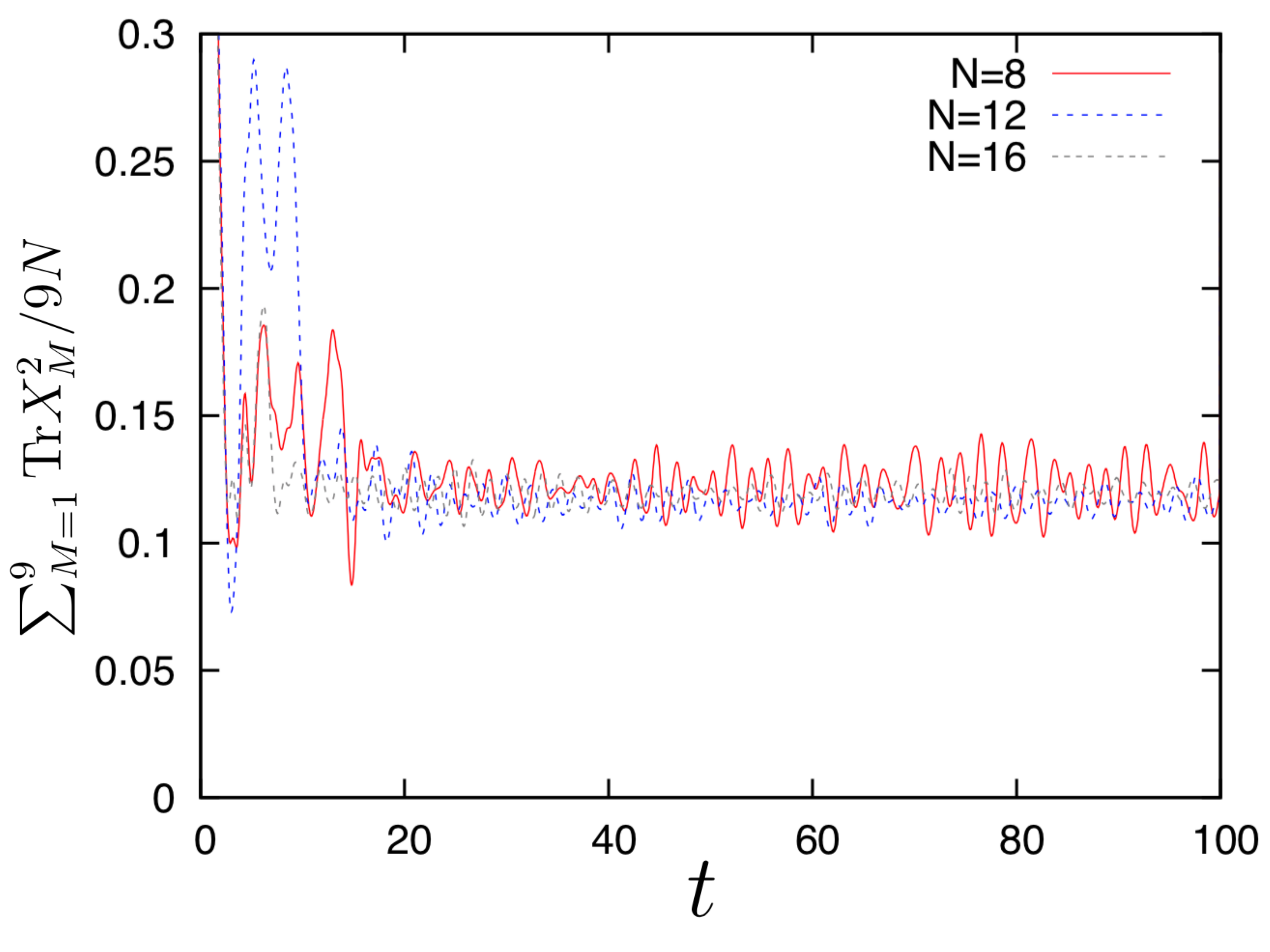}}
\caption{$\sum_{M=1}^9{\rm Tr}X_M^2/9N$ for $N=8,12$ and $16$, for $L=5.0$, $V=0$ and 
 $\sqrt{N}\sigma=0.12$. The bottom panel is a zoom-up of the top one.
 Qualitatively similar behaviours can be seen and the fluctuation at late-time becomes smaller as $N$ becomes larger. 
 }\label{L5V0O12trX}
\end{figure}

\begin{figure}[htb]
\begin{center}
\scalebox{0.35}{
\includegraphics{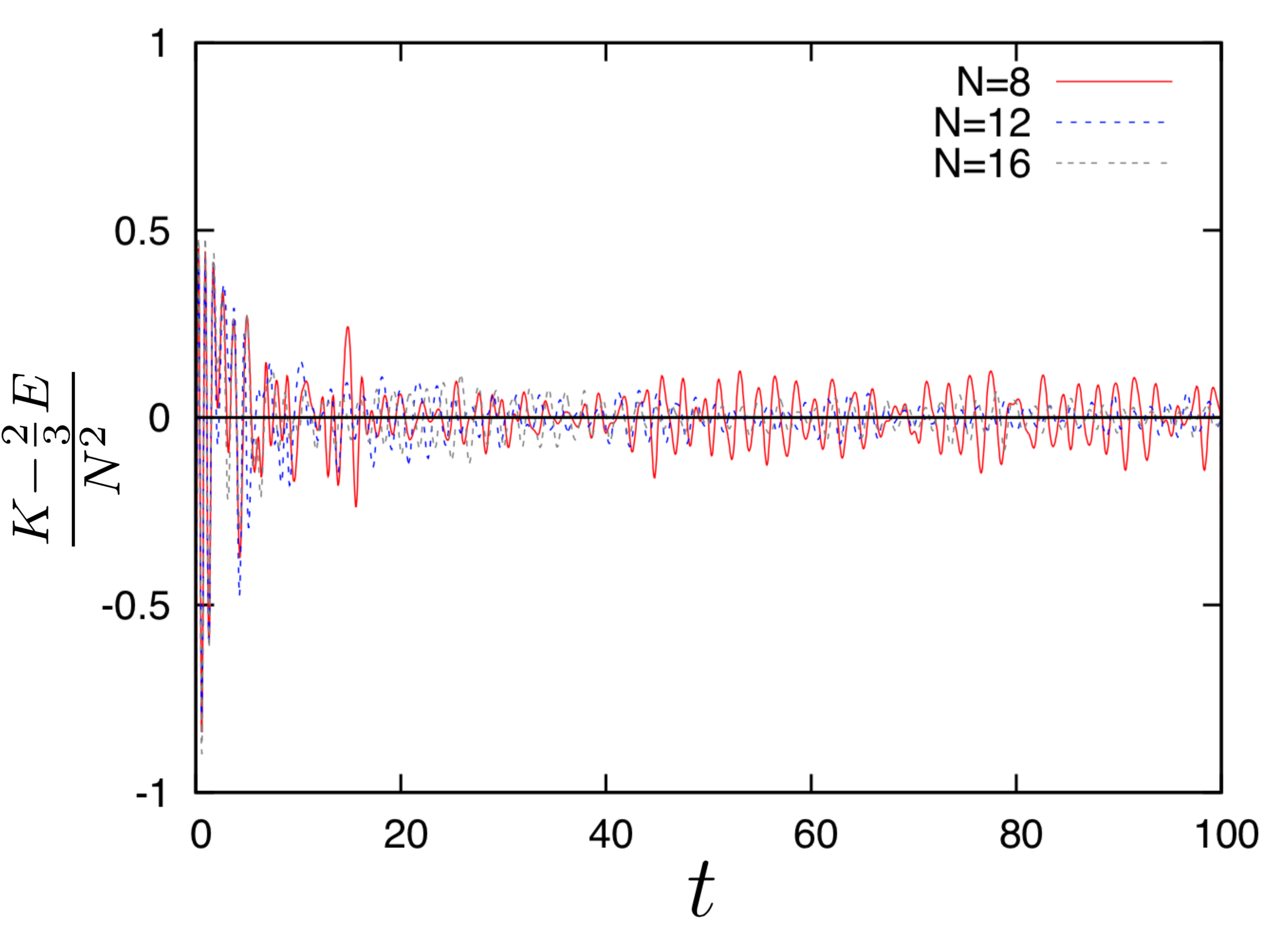}}
\end{center}
\caption{
$\frac{K - \frac{2}{3}E}{N^2}$ for $N=8,12$ and $16$, for $L=5.0$, $V=0$ and $\sqrt{N}\sigma=0.12$. 
Here $K$ and $E$ are the kinetic energy and total energy. When the bound state is formed, the long-time average 
of $\frac{K - \frac{2}{3}E}{N^2}$ must be zero due to the virial theorem. The fluctuation around zero becomes smaller 
as $N$ becomes larger.  
}\label{K-avK_L5V0O12}
\end{figure}

In Fig.~\ref{L5V0O12trX}, we plot ${\rm Tr}X_1^2/N$ for $N=8,12,16$. 
We can see qualitatively similar behaviours, and the fluctuation at late-time becomes smaller 
as $N$ becomes larger. 
When a bound state is formed, the Virial theorem relates the kinetic energy $K$ and the potential energy $V$ as 
$\langle K\rangle =2\langle V\rangle=\frac{2}{3}E$, where $\langle\ \cdot\ \rangle$ stands for the time average 
and $E=K+V$ is the total energy, which is conserved. Therefore, $({K - \frac{2}{3}E})/{N^2}$ is a good 
measure for the fluctuation. As we can see from Fig.~\ref{K-avK_L5V0O12}, this quantity fluctuates around zero 
and suppressed more as $N$ becomes larger. 

\begin{figure}[htbt]
\centering  
\scalebox{1.34}{
\subfloat{\includegraphics[clip, width=2.2in]{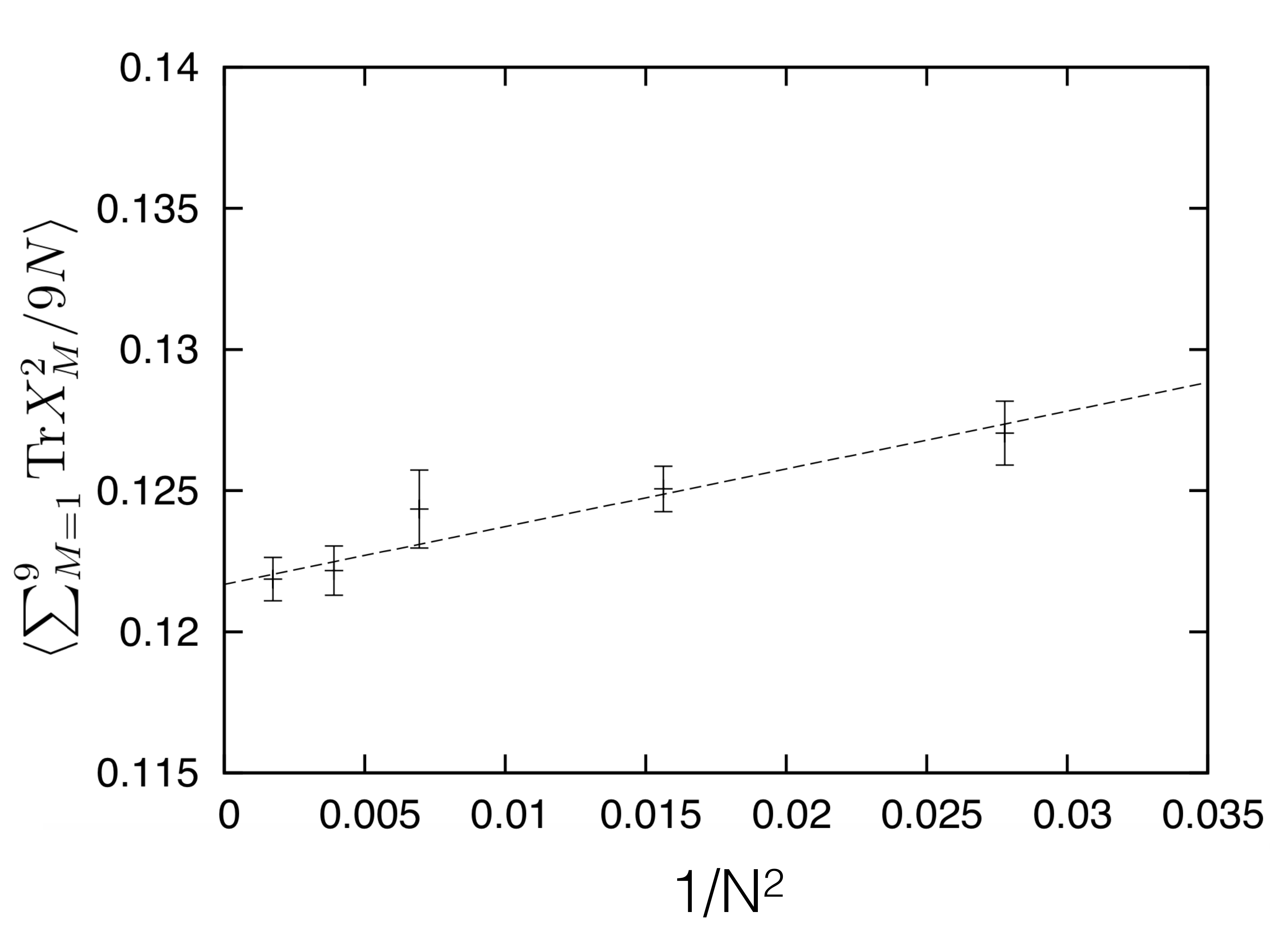}}}
\\
\scalebox{1.34}{
 \subfloat{\includegraphics[clip, width=2.2in]{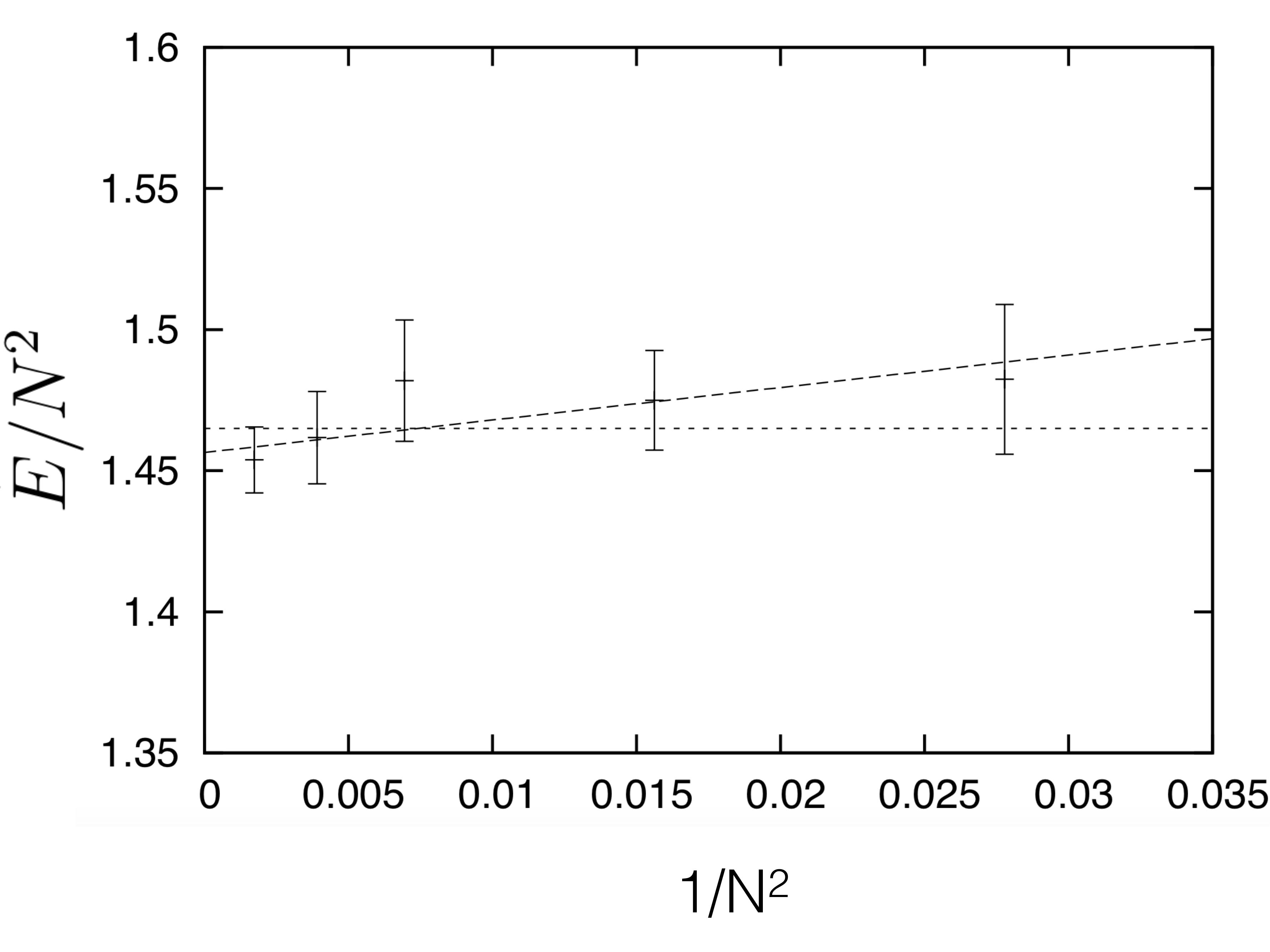}}}
\\
\scalebox{1.34}{
 \subfloat{\includegraphics[clip, width=2.3in]{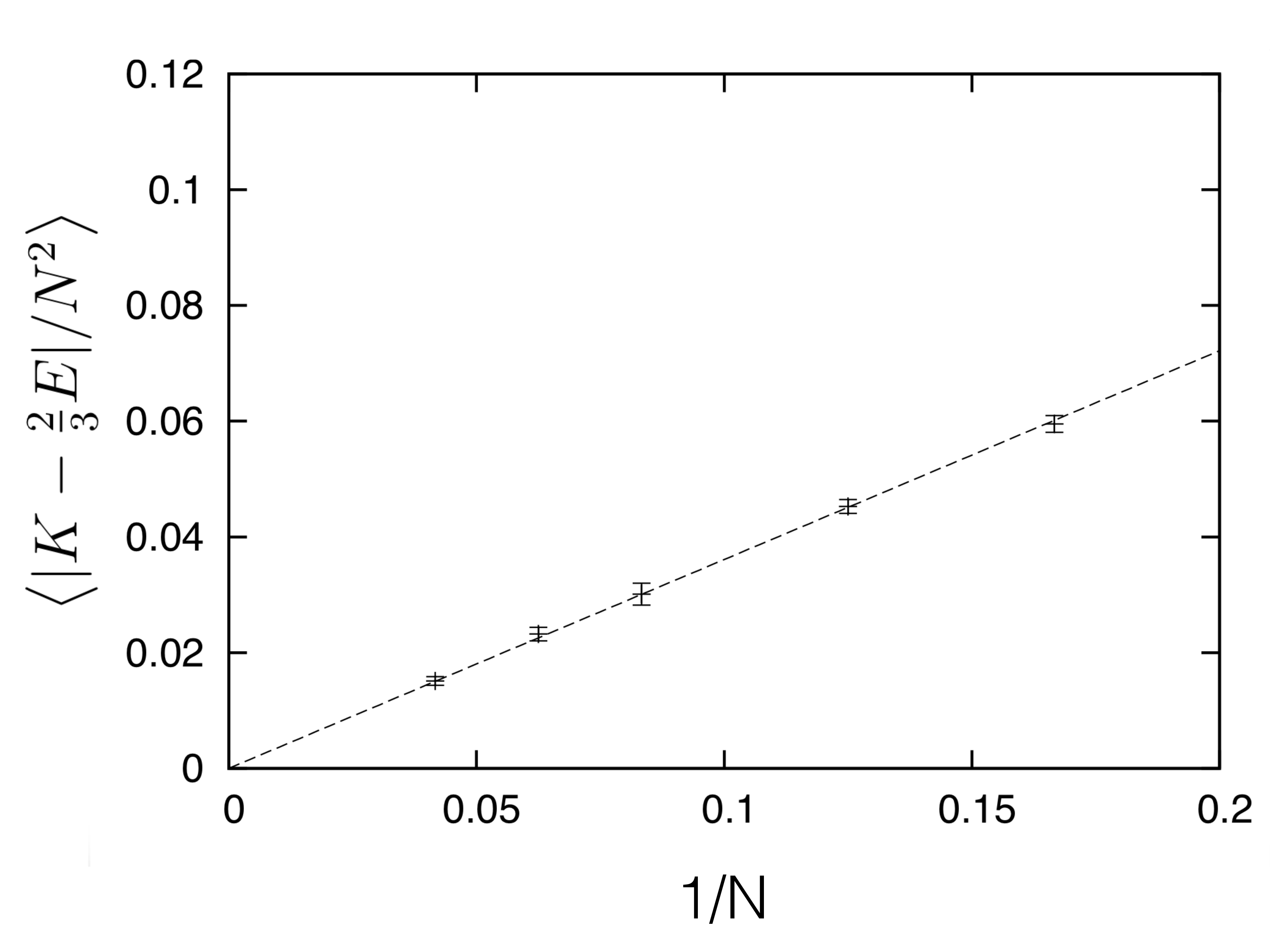}}}
\caption{$N$-dependence of $\langle \sum_{M=1}^9{\rm Tr}X_M^2/9N\rangle$ (top), 
 $E/N^2$ (middle) and $\left\langle |K - \frac{2}{3}E|/N^2\right\rangle$ (bottom). 
 Collision parameters are $L=5.0$, $V=0$ and $\sqrt{N}\sigma=0.12$. }\label{L5V0O12_various_N}
\end{figure}

In order to see the statistical property at late-time, we take the time average at $50\le t\le 100$ 
and then take an  average over random initial configurations (50 samples for $N=4,8$, 20 samples for $N=12$, 
15 samples for $N=16$ and 10 samples for $N=24$). 
The error bar is estimated purely statistically. 
The results are plotted in Fig.~\ref{L5V0O12_various_N}, especially 
the bottom figure of Fig.~\ref{L5V0O12_various_N} shows how $N$ dependence 
appears in  $\langle |K - \frac{2}{3} E| \rangle$. 
Note that since $|K - \frac{2}{3} E|$ is an absolute value, 
its fluctuation is always added up. The fact that 
\bea
\lim_{N \to \infty} \langle |K - \frac{2}{3} E| \rangle = 0
\eea
suggests that 
the fluctuation is suppressed as $N$ becomes larger. 
In finite $N$, 
$\left\langle |K - \frac{2}{3}E|/N^2\right\rangle$ is proportional to $1/N$, 
while the corrections to $E/N^2$ and $\langle \sum_{M=1}^9{\rm Tr}X_M^2/9N\rangle$ are proportional to $1/N^2$. 
$E/N^2$ is consistent with 
$c_1 + c_2/N^2$ behaviour with some constant $c_1$ and $c_2$, where $c_2$ could be zero.  


\subsubsection{Large-$N$ limit with fixed $E/N^2$}

\begin{figure}[htbt]
\centering  
\scalebox{1.3}{
\subfloat{\includegraphics[clip, width=3.0in]{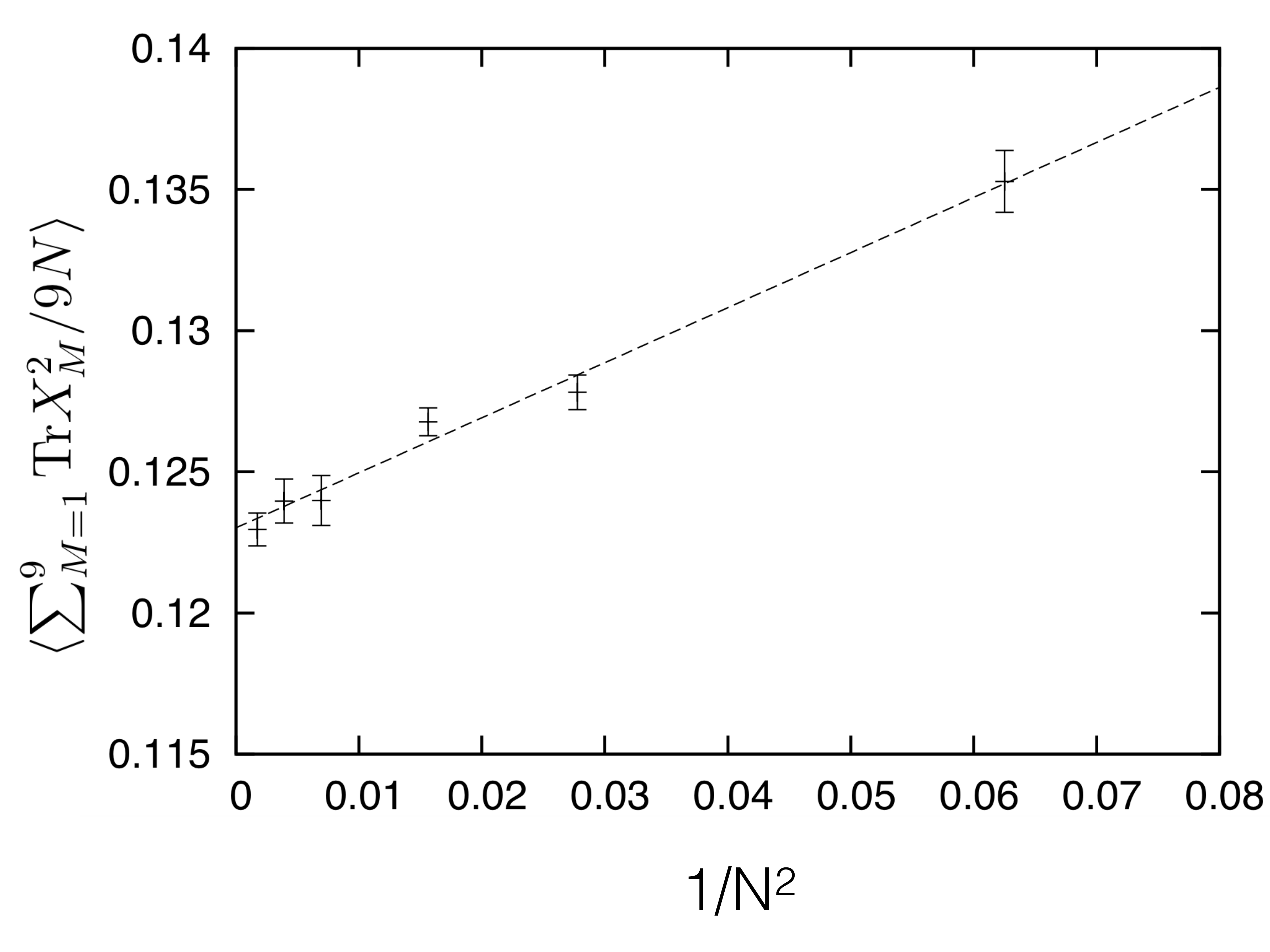}}}
\\
\scalebox{1.3}{
 \subfloat{\includegraphics[clip, width=3.0in]{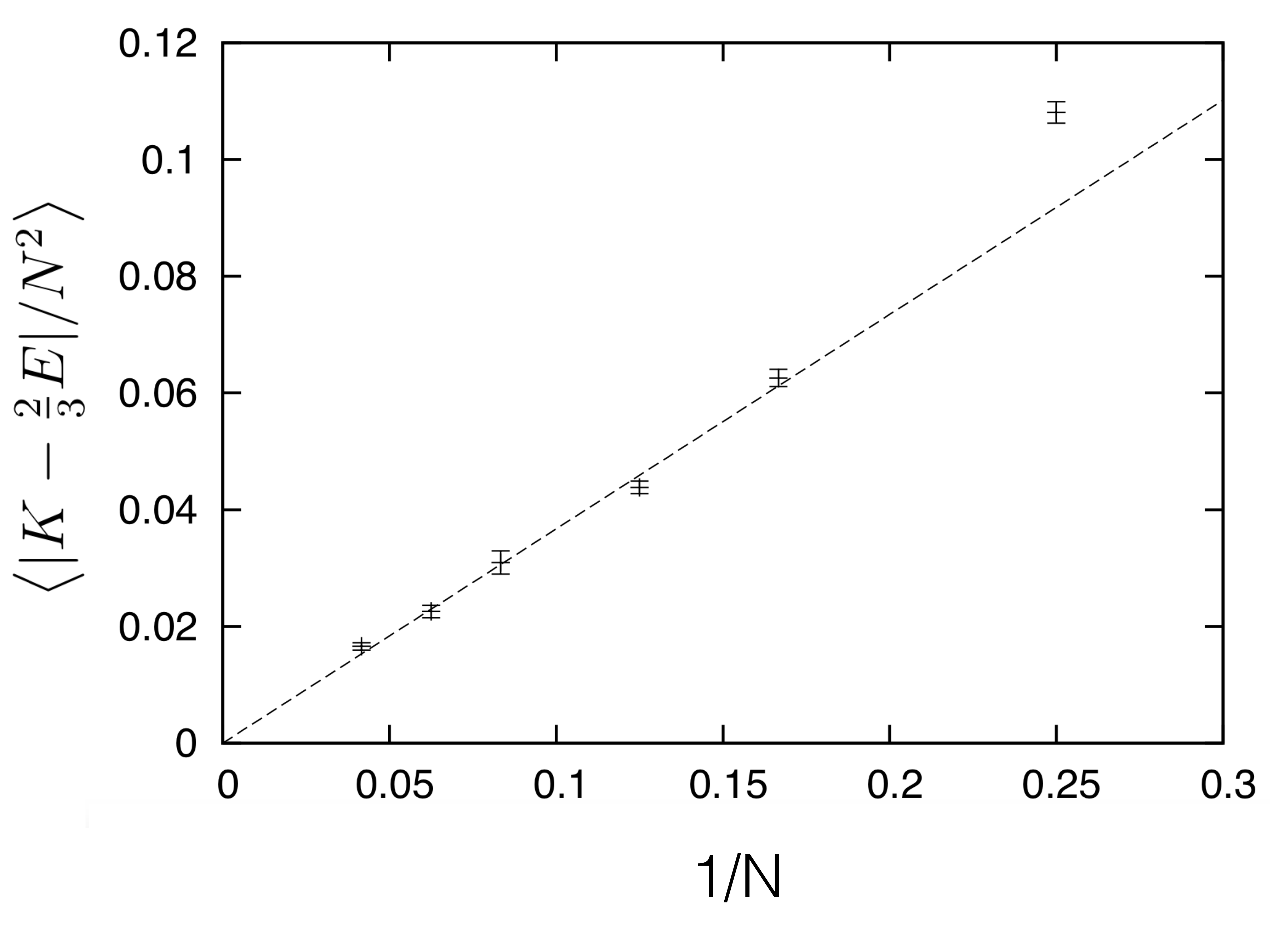}}}
\caption{$N$-dependence of $\langle \sum_{M=1}^9{\rm Tr}X_M^2/9N\rangle$ (top) and 
$\left\langle |K - \frac{2}{3}E|/N^2\right\rangle$ (bottom), where the energy at each run is fixed to $E/N^2=1.5$. 
$\langle \sum_{M=1}^9{\rm Tr}X_M^2/9N\rangle$ behaves as ${\rm const}.+{\rm const}./N^2$, 
while $\left\langle |K - \frac{2}{3}E|/N^2\right\rangle$ is consistent with ${\rm const.}/N$ behaviour.  
}\label{L5V0O12E15_various_N}
\end{figure}

Let us consider another, perhaps more natural large-$N$ limit, in which $E/N^2$ is fixed exactly\footnote{In the large $N$ limit, which corresponds to the thermodynamic limit, this is the same as fixing the temperature of the system.}.  
We first generate initial configurations with $L=5.0$, $V=0$ and $\sqrt{N}\sigma=0.12$, 
and then set $E/N^2$ to $1.5$, by rescaling scalars as we did in the end of \S~\ref{sec:bunch_formation}. 
We collected 50 samples for $N=4,8$, 20 samples for $N=12$, 15 samples for $N=16$ and 10 samples for $N=24$. 
We use $50\le t\le 100$ again. 
In principle, since the energy is fixed, the time average for all samples with different initial configurations should give the same average value within errors for a sufficiently long time interval. 
(Still, we terminate the simulation at $t=100$ 
so that the conservation of the energy is not violated due to the discretization error.)
Therefore, statistical  fluctuations of the average $\langle \sum_{M=1}^9{\rm Tr}X_M^2/9N\rangle$  become much smaller compared to Fig.~\ref{L5V0O12_various_N}.  
The results are plotted in Fig.~\ref{L5V0O12E15_various_N}. We can see that 
the correction to $\langle \sum_{M=1}^9{\rm Tr}X_M^2/9N\rangle$ also starts with $1/N^2$.

\subsubsection{Large-$N$ limit and thermalization}

As we have seen, the system shows typical thermalization process and the fluctuation of macroscopic quantities disappears at large $N$. 
Therefore, the macroscopic quantities converge to the same value at late time, irrespectively of the initial condition, as long as the energy is the same. The information of the initial condition is hidden in the $1/N$ correction. This is exactly what we expect. Since the large $N$ limit corresponds to the thermodynamic limit 
there is no distinguish between micro canonical ensemble (this is what we did) and canonical ensemble where temperature is fixed. The difference only appears in $1/N$ expansion. Note that an exact thermalization occurs only in the large $N$ limit.

\subsection{Thermalization time}\label{sec:timescalesection}
\begin{figure}[htb]
\begin{center}
\scalebox{0.35}{
\includegraphics{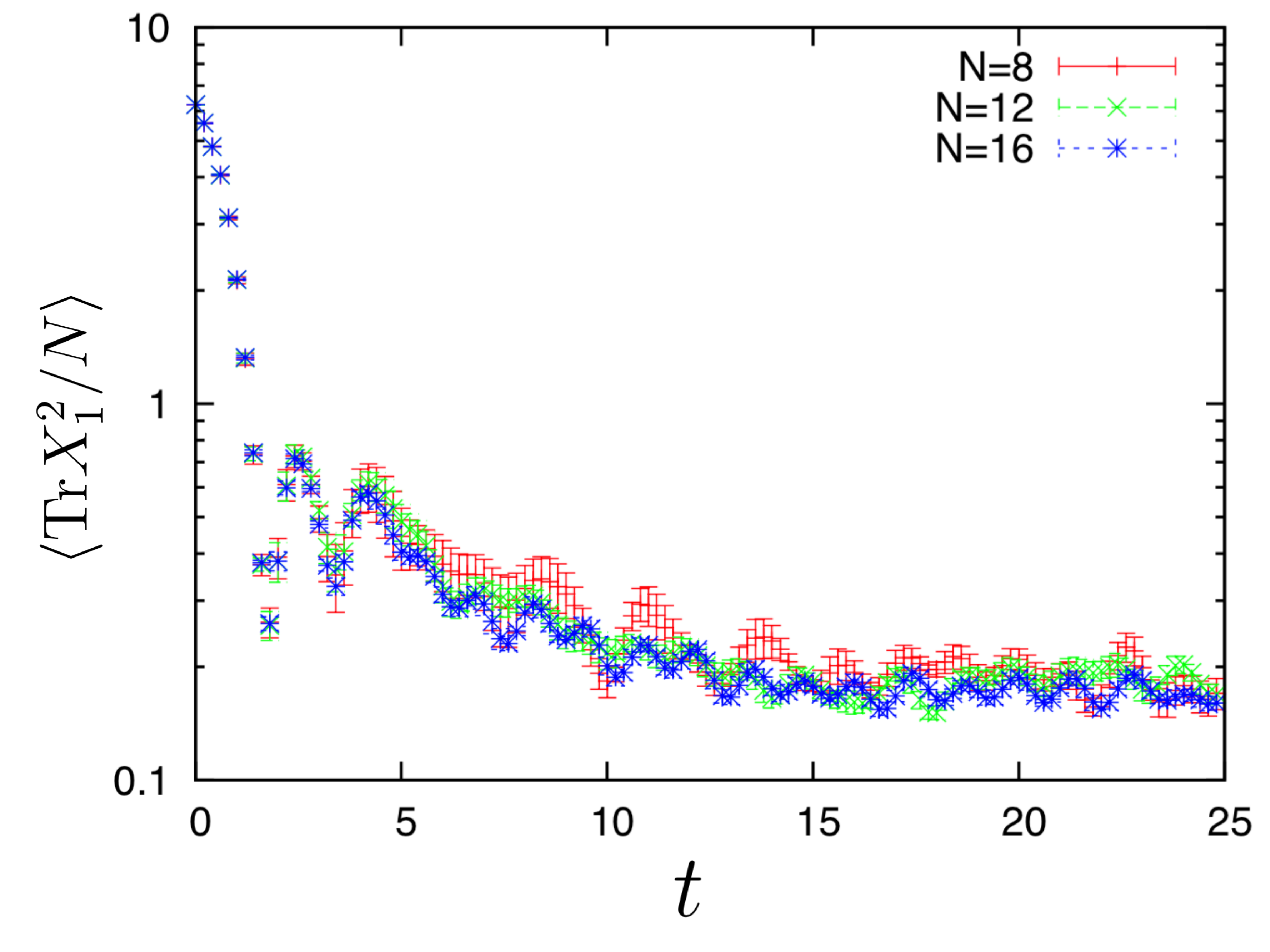}}
\end{center}
\caption{
One-point function 
$\langle{\rm Tr}X_1^2/N\rangle$ for $N=8,12,16$, $L=5.0$, $V=1.0$ and $\sqrt{N}\sigma=0.16$. 
Average over 10 samples.  
Results with different values of $N$ are not distinguishable within statistical error. 
}\label{TrX_av_10samples}
\end{figure}

Let us investigate the thermalization time. 
There are several different ways to define the thermalization time.  
Although it involves the dual interpretation, perhaps one of the most natural ones is the time scale for the ``black hole'' formation. 
In Fig.~\ref{TrX_av_10samples}, we plot the history of $\langle{\rm Tr}X_1^2/N\rangle$ for $N=8,12,16$, $L=5.0$, $V=1.0$ and $\sqrt{N}\sigma=0.16$. 
We used 10 different sets of random numbers and took average. Data at $N=12$ and $N=16$ are indistinguishable within errors, suggesting that the time evolution of $\langle{\rm Tr}X_1^2/N\rangle$ has a well-defined large-$N$ limit. 
Remember that large $N$ limit is a thermodynamic limit, where canonical ensemble and microcanonical ensemble are distinguishable only by looking at the $1/N$ suppressed order.  
Therefore, we call the system get thermalized at the late-time 
if the system becomes indistinguishable 
``typical states'', and the deviations for the macroscopic quantities from the typical states are only of order $1/N$. 
Then, to read-off the thermalization time, {\it i.e.,} the time-scale system get thermalized, 
we have to read-off the decay 
of the deviation at late-time in great detail to the order of $1/N$. 
However this is hard generically since the error bar is too large and we have initial-condition-dependence for the late-time average. 

\begin{figure}[htb]
\begin{center}
\scalebox{0.35}{
\includegraphics{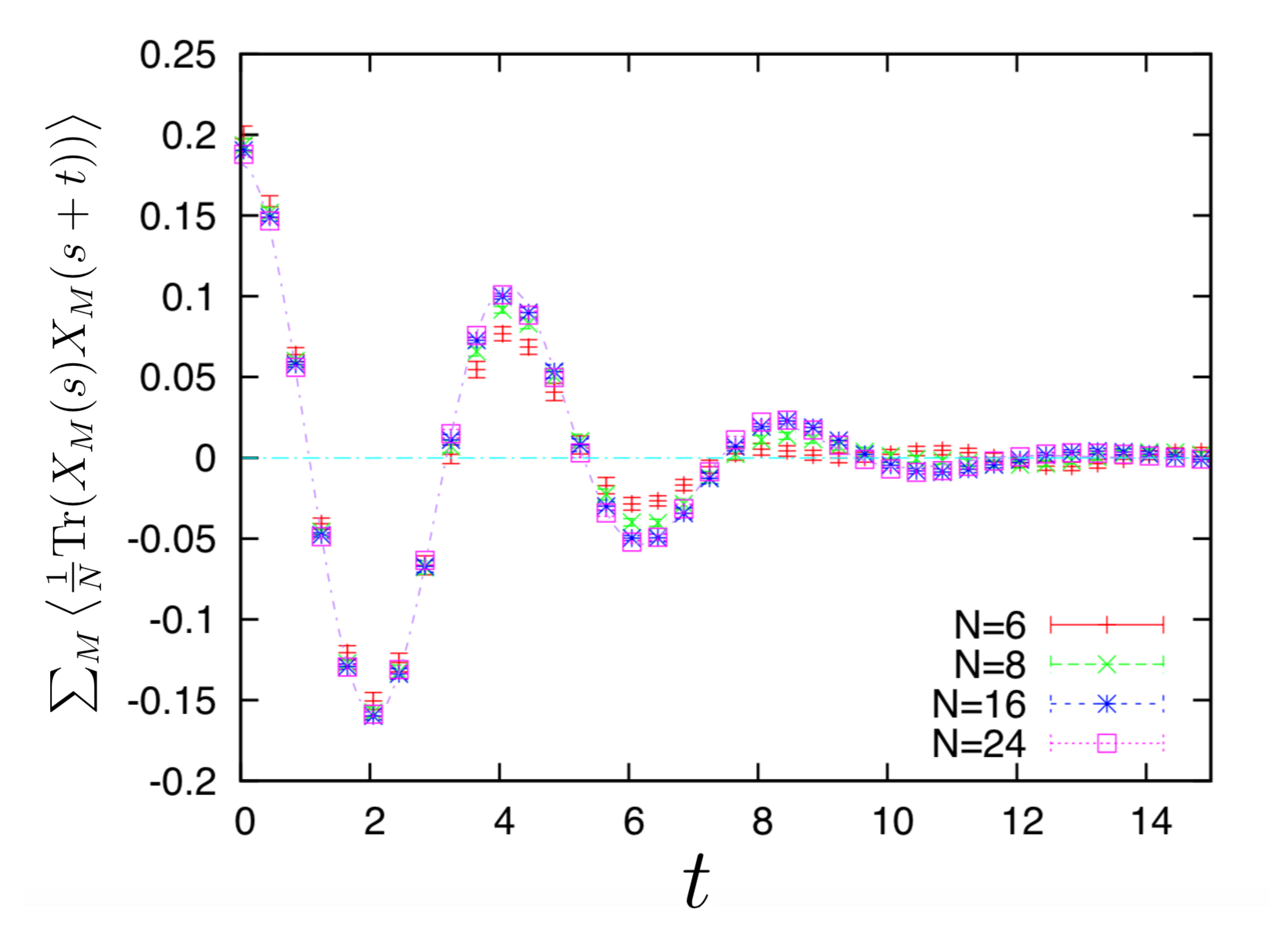}}
\end{center}
\caption{
$f(t)\equiv
\sum_M\left\langle\frac{1}{N}{\rm Tr}(X_M(s)X_M(s+t))\right\rangle$ for $N=6,8,16$, and $24$, $L=5.0$, $V=0$ and $\sqrt{N}\sigma=0.16$. 
The fitting line is for $N=24$, with the ansatz $f(t)\simeq c \, e^{-\Gamma t^2}\cos\left(\omega t\right)$ 
and fit parameters $c=0.182$, $\Gamma=0.0311$ and $\omega=1.48$. 
Better convergence to $N=\infty$ can be seen at earlier time. 
}\label{2pt_v1}
\end{figure}

To avoid this issue, we define the thermalization time by another way. 
This is defined by the time scale for a small perturbation added on top of the bound state to become invisible. 
Note that if $N$ is not sufficiently large, then the thermalization time defined in this way would depend on the detail of the perturbation.
Let us consider two-point function along time direction 
which does not require the perturbation, and from that, we will read off the  
thermalization time.  
Let us consider a gauge-invariant operator
\begin{eqnarray}
\frac{1}{N}{\rm Tr}(X_M(s)W_{s,s+t}X_M(s+t)W_{s,s+t}^\dagger), 
\end{eqnarray}
where $W_{s,s+t}$ is the Wilson line, 
\begin{eqnarray}
W_{s,s+t}={\rm P}e^{i\int_s^{s+t}dt'A(t')}. 
\end{eqnarray}
In our setup we took the $A=0$ gauge, so 
\begin{eqnarray}
\frac{1}{N}{\rm Tr}(X_M(s)W_{s,s+t}X_M(s+t)W_{s,s+t}^\dagger)
=
\frac{1}{N}{\rm Tr}(X_M(s)X_M(s+t)). 
\end{eqnarray}
Let us consider 
\begin{eqnarray}
f(t)
\equiv
\sum_M\left\langle\frac{1}{N}{\rm Tr}(X_M(s)X_M(s+t))\right\rangle,  
\end{eqnarray}
where the time average $\langle\ \cdot\ \rangle$ is taken at late-time. We have already seen that $f(0)$ is of order one. 
When $t$ becomes large, $f(t)$ decays because $X_M(s+t)$ ``forgets'' the information of $X_M(s)$. 
Therefore, it should be possible to determine the thermalization time from the time scale for the decay of $f(t)$. 
The numerical result is shown in Fig.~\ref{2pt_v1}. 
It turned out that $f(t)$ can be fitted well as   
\begin{eqnarray}
f(t)\simeq c \, e^{-\Gamma t^2}\cos\left(\omega t\right), 
\end{eqnarray}
where the overall constant $c=\sum_M \left\langle\frac{1}{N}  {\rm Tr}X_M(t)^2\right\rangle$, the decay width $\Gamma$ and the frequency $\omega$ are of order $N^0$. 
The width $\Gamma$ naturally gives the time scale for the thermalization, 
\begin{eqnarray}
\label{ttestimates1}
{\rm (Thermalization \, time)} \sim 1/\sqrt{\Gamma}\sim N^0. 
\end{eqnarray}

\begin{figure}[htb]
\begin{center} 
\scalebox{0.35}{
\includegraphics{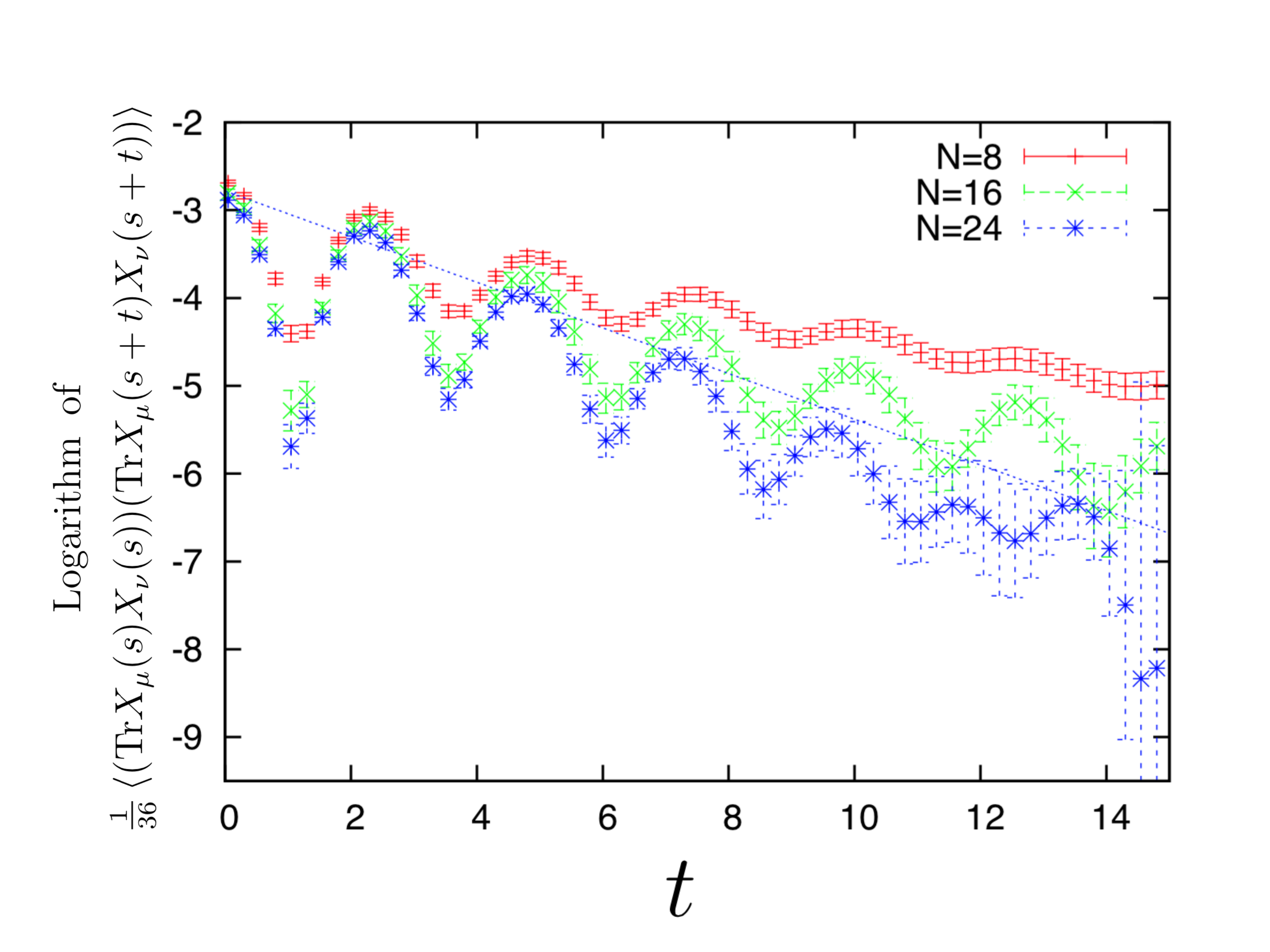}}
\end{center}
\caption{
 Logarithm of 
$\frac{1}{36}\left\langle({\rm Tr}X_\mu(s)X_\nu(s))({\rm Tr}X_\mu(s+t)X_\nu(s+t))\right\rangle$ as a function of $t$, 
for $N=8,16$, and $24$, $L=5.0$, $V=0$ and $\sqrt{N}\sigma=0.16$. 
The fitting line is for $N=24$, with $0.062\cdot e^{-0.245t}$. 
Better convergence to $N=\infty$ can be seen at earlier time. 
}\label{2pt_v2}
\end{figure}

We can also consider connected two-point functions of gauge invariant operators. 
We consider one of the operators studied by Berenstein and collaborators, 
${\rm Tr}(X_M(t)X_N(t))$ $(M \neq N)$.  
Although we did not find a simple fitting function for this, the decay is consistent with $e^{-\Gamma' t}$ (Fig.~\ref{2pt_v2}). 
Then the thermalization time defined by this operator is 
\begin{eqnarray}
{\rm (Thermalization \, time)} \sim 1/\Gamma' \sim N^0. 
\end{eqnarray}
Fig.~\ref{2pt_v2} suggests that the thermalization time, defined from the exponential decay, converges to finite value in the large $N$ limit. 
This is quite interesting since it is consistent with finiteness of quasinormal modes in the black hole background,  
although this convergence is more subtle at late time due to finite $N$ effect.  
As is discussed in \cite{Iizuka:2008eb}, although the correlation function keep decaying to zero at any finite order in the $1/N$-expansion, 
the decay should disappear once full finite-$N$ corrections are taken into account. 
The fat tail at long time scale in Fig.~\ref{2pt_v2} would demonstrate this property.

\subsection{Under what initial conditions, does thermalization occur?}\label{sec:when_thermal}
Let us consider at what values of $V$ and $\sigma$ 
the thermalization, or equivalently the formation of the bound state, can be realized. 
For simplicity we fix $L$ to be $5.0$. 
Firstly let us note that the off-diagonal elements are the source for the non-linearity, which is responsible for the formation of the bound state. 
If we set the off-diagonal elements to be zero ($\sigma=0$) at $t=0$, then the off-diagonal elements are not generated by the classical equations of motion and  
no interaction appears at all. In such case, two bunches just pass through each other. 

Even if $\sigma$ is nonzero, if it is too small, the non-linearity 
cannot grow large enough before the collision. 
Therefore, it is clear that $\sigma$ must be sufficiently large for the bound state formation.  
However, as we will see shortly, it turns out that $\sigma$ should not be too small but also must not be too large in order to form the bound state. 
This indicates that  if  the $\sigma$ accelerates the two bunches before the first collision too much, they tend to pass by without forming the bound state.

\begin{figure}[htb]
\begin{center}
\scalebox{0.35}{
\includegraphics{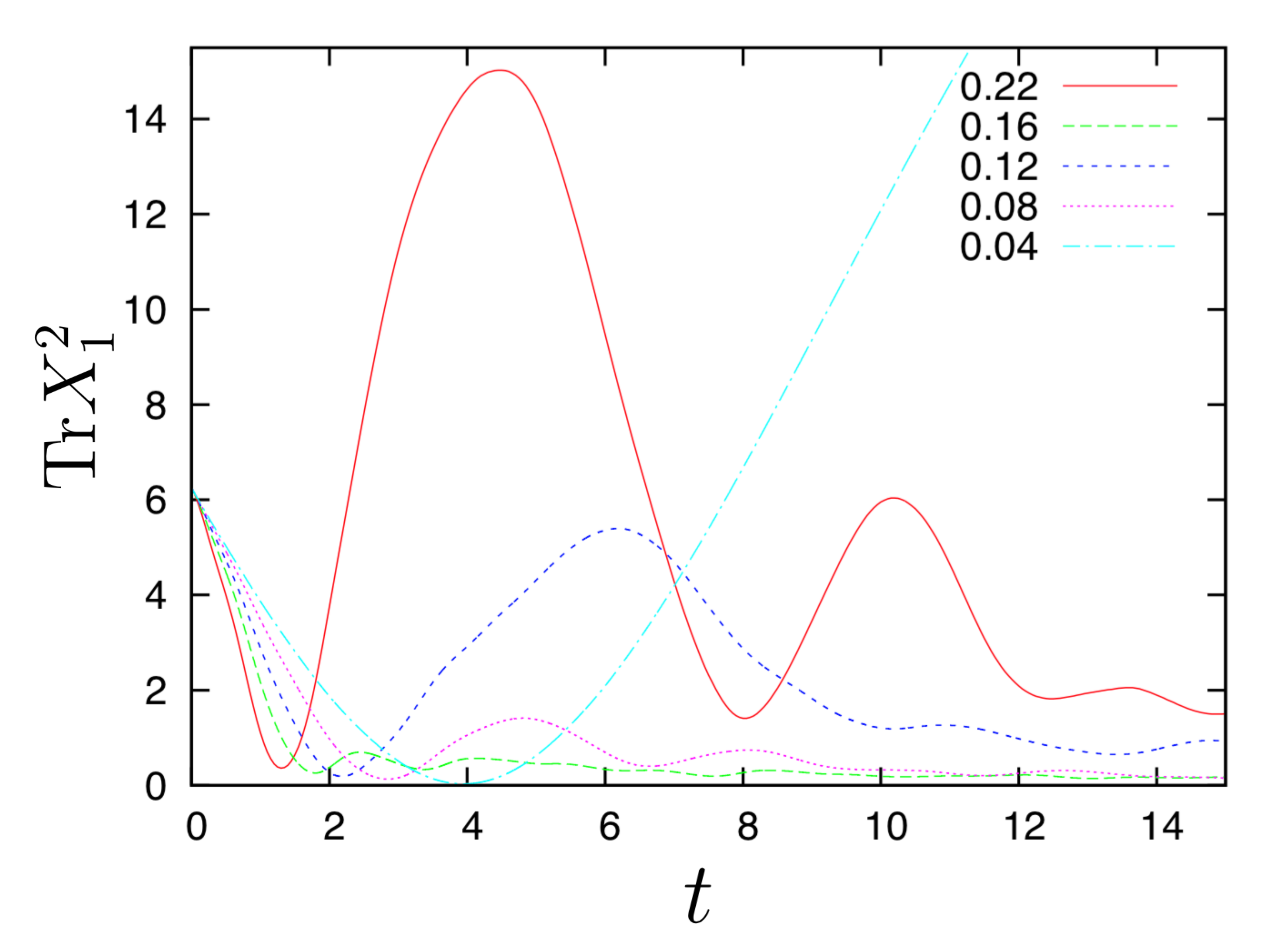}}
\end{center}
\caption{
${\rm Tr} X_1^2$ for $N=16$, for $L=5.0$, $V=1.0$ and various values of $\sqrt{N}\sigma$. 
}\label{trX1_bound_or_unbound}
\end{figure}

In order to illustrate these, we show the evolution of ${\rm Tr}X_1^2/N$ for $N=16$, for $L=5.0$, $V=1.0$ and various values of $\sqrt{N}\sigma$ 
in Fig.~\ref{trX1_bound_or_unbound}. 
For $\sqrt{N}\sigma=0.16, 0.12$ and $0.08$, the late-time behaviours at $t\gtrsim 20$ look qualitatively the same. 
A fact that a bounce at $\sqrt{N}\sigma=0.12$ is larger than those at $\sqrt{N}\sigma=0.08$ and $\sqrt{N}\sigma=0.16$ suggests very complicated dynamics. 
At $\sqrt{N}\sigma=0.04$, on the other hand, two bunches just pass through with each other. 
It might be possible that the bunches comes back later after very long time and form a bound state; 
in fact sometimes we observe that ${\rm Tr}X_1^2/N$ becomes 
as large as $O(100)$ and then come back and form a single bound state.  
As we can see from the early-time behavior in Fig.~\ref{trX1_bound_or_unbound}, 
as $\sigma$ increases the bunches are accelerated more before the collision (i.e. ${\rm Tr}X_1^2/N$ decreases more rapidly). 
So whether a single bound state corresponding to thermalised state is formed or not 
depends on a very subtle competition of the acceleration before the collision and 
deceleration after the collision, both of which are enhanced by a larger $\sigma$.  

\begin{figure}[htb]
\begin{center}
\scalebox{0.35}{
\includegraphics{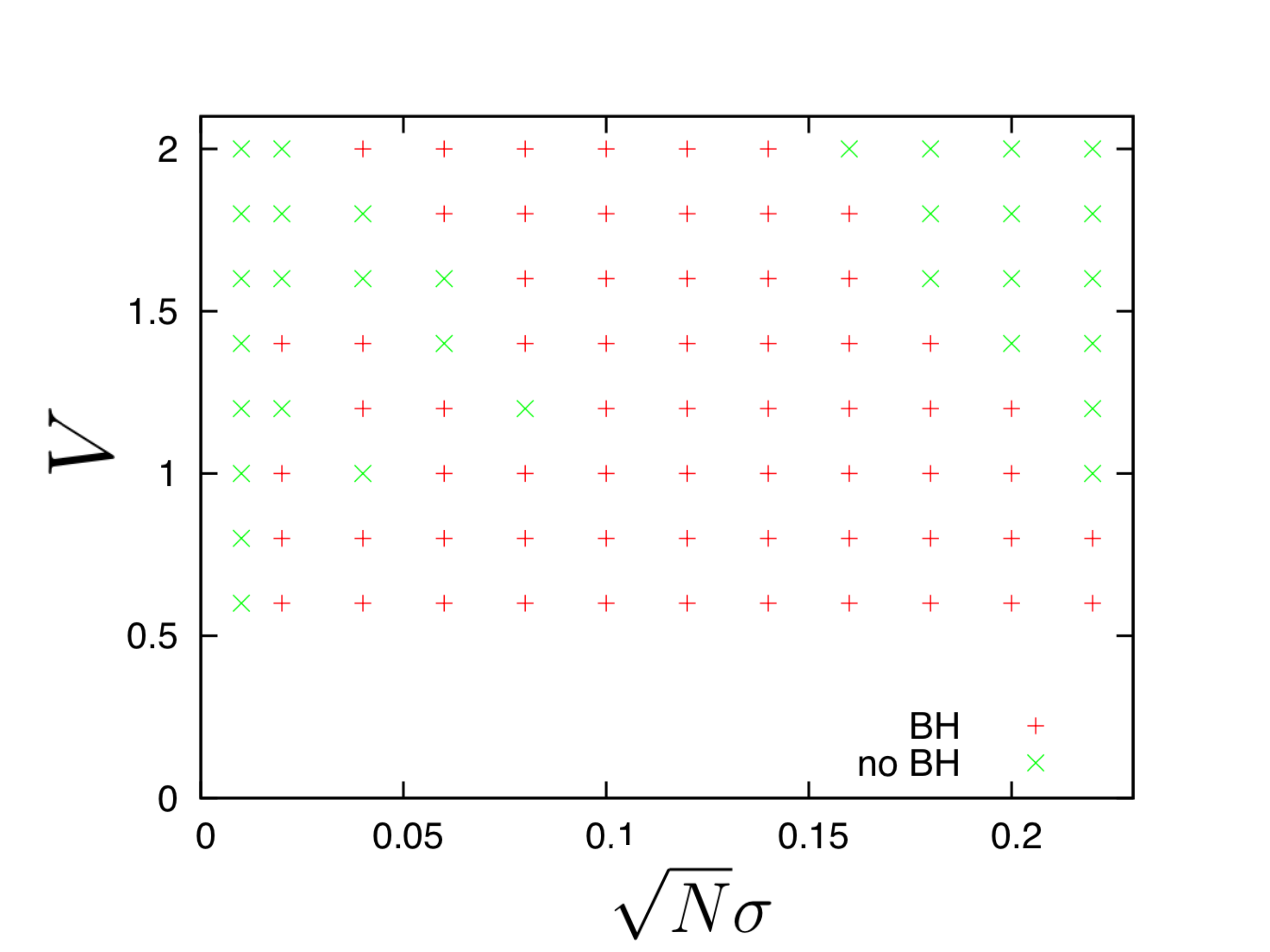}}
\end{center}
\caption{
 $N=16$, $L=5.0$. 
 The green ``$\times$" symbols show $(V,\sqrt{N}\sigma)$ in which two bunches passed through with each other 
and ${\rm Tr}X_1^2/N$ became at least twice bigger than the initial value. A single-bunch state would be formed after long time, but not immediately. 
The red ``$+$" symbols show $(V,\sqrt{N}\sigma)$ where a single bound state is formed without bouncing too much.  
We used the same set of random numbers $a_\mu^{ij}$ and $b_\mu^{ij}$ for all $V$ and $\sqrt{N}\sigma$. 
}\label{BH_or_not}
\end{figure}

The green ``$\times$" symbols in Fig.~\ref{BH_or_not} show $(V,\sqrt{N}\sigma)$ in which two bunches passed through with each other 
and ${\rm Tr}X_1^2/N$ became at least twice bigger than the initial value. A single-bunch state would be formed after long time, but not immediately. (See e.g. $\sqrt{N}\sigma=0.22$ in Fig.~\ref{trX1_bound_or_unbound}.) 
The red ``$+$" symbols show $(V,\sqrt{N}\sigma)$ where a single bound state is formed without bouncing too much.  
We can see that a single bound state is not formed when $\sqrt{N}\sigma$ is too large, which means the acceleration before the collision 
beats the deceleration after the collision.   
We cannot get any conclusive pattern from these, 
probably the result depends on the choice of the random numbers. We left this issue  
as open questions.

\section{Summary and discussions}

In this paper, we considered the head-on collisions of two bunches of D0-branes and study its real-time evolution in the BFSS matrix model. 
Especially our interests in this study are
how time evolution differs and under what conditions 
the big bound states corresponding thermalized state are formed and what is the its time-scale. 

We have seen that all of ${\rm Tr}X_M^2/N$ where $M = 1, \cdots , 9$ converges at late time to the 
same values and rotational symmetry is restored irrespective to the anisotropic initial conditions in \S \ref{sec:bunch_formation}. Especially the insensitivity to the initial conditions strongly suggest the thermalization and formation of a large bound state dual to quantum black holes in late time.  In \S \ref{sec:Virialand1/N}, for various scaling limit, we have checked that at the thermalized stage, Virial theorem, which is expected to hold at the equilibrium state, is in fact satisfied and deviation from that occurs only at the subleasing order in $1/N$ expansion. 
In \S \ref{sec:timescalesection}, we estimated thermalization time from the time-direction two point functions and its exponential decay at late time, and shows that the timescale is $N^0$ order, which seems to converge in the large $N$ limit. This result is very interesting since it is consistent with quasi-normal mode in the dual black hole.
Finally in \S \ref{sec:when_thermal}, we studied the initial condition dependence for the thermalization.  Clearly for small $\sigma$, the off-diagonal sources for non-commutativity, two D0-brane bunches pass by while we notice that for too large $\sigma$, two bunches also pass by in such cases. We
however have no clear interpretation for this results.

Our analysis is conducted under several assumptions. The big assumptions 
we take in our analysis is that we can completely neglect the quantum effects and fermion effects. 
Quantum effects are important as the effective coupling 
becomes larger, and fermions are also very important since they change the interactions 
between two bunches of D0-branes. It is quite possible that once we take into account these effects, then the time-scale we estimated at \S \ref{sec:timescalesection} can be modified significantly. Obviously more detail study is necessary to get conclusive results.

To better understand under what initial conditions does the system thermalized, let 
us discuss what is the typical states in the BFSS matrix model. 
If the final states is made up by sets of almost commuting matrices, 
then the entropy of the gas is made up by $N$ D0-branes, which has only $O(N)$ entropy. 
Off-diagonal elements are highly suppressed in such case and does not contribute to the entropy. 
On the other hand, a single large black hole, 
which is a single bound state of the D0-branes, has the entropy of order $N^2$,  
because all of the off-diagonal elements are excited. 
We can also consider multi-black hole states. 
Clearly a single black hole is entropically and therefore free-energy-wise dominant, since 
all of the off-diagonal elements are equally excited and contributes to the entropy.

However there are subtle issues. 
In the BFSS matrix model, because moduli space is not bounded, 
when the two bunches pass through each other, it is possible, at least classically, that 
they just pass by and run to infinity. 
This is all due to the fact that all of the adjoint scalars have zero mass\footnote{
There is a subtlety if the moduli space is infinity, how the ergodicity can hold. 
Probably the appropriate limit is that, first we introduce a proper IR cutoff for matrix eigenvalues (for examples, by putting D0-branes 
in a box or by introducing small but nonzero mass for the adjoint matrices) 
so that the phase space becomes bounded, 
the ergodicity can hold for generic initial conditions. And then we evaluate 
spectrum and correlators and in the end, we remove the IR cut-off. 
For the BMN matrix model \cite{Berenstein:2002jq}, this IR problem is avoided.}.
However by taking into account quantum fluctuations, this is unlikely, because in the real system involving quantum fluctuation, 
there are always small but nonzero quantum fluctuation and when the non-linearity from the interaction is large enough at late time, two bunches are likely to be pulled each other and merge to a single bound state. 
If the entropy of the black hole is large enough ({\it i.e.} $N$ is large enough), 
then it is unlikely to see a large deviation from the black hole 
within a finite simulation time. 

Note that since our calculation is classical, it is crucial to take small effective coupling limit {\it i.e.,} $\lambda_{eff} \to 0$ to justify our results. 
Even though effecting coupling is small, 
the equation of motion can accumulate large non-linearity at late time,
which can capture non-perturbative soliton formation physics. 
A large $N$ effect (which plays the same role as thermodynamic limit) and the late time universality effects (which capture enough non-linearity of the interaction and scramble) are crucial ingredients for the thermalization.  
There is an argument that the classical time evolution of the BFSS matrix model is stochastic \cite{Savvidy:1982jk}. Such natures are probably crucial for the system get thermalized 
and show the universal behaviour at late time.

\section*{Acknowledgement} 
\hspace{0.51cm}
The authors would like to thank D.~Berenstein, G.~Gur-Ari, G.~Ishiki, D.~Kabat, J.~Maltz, S.~Shenker, M.~Shigemori, H.~Shimada, 
S.~Shimasaki, S.~Susskind, and S.~Yamaguchi for stimulating discussions and comments.
S.A. is supported in part by the Grant-in-Aid of the Ministry of Education, Science and Technology, Sports and Culture
(No. 25287046) and the Strategic program for Innovative Research (SPIRE) Field 5, and JICFuS \cite{JICFUS}.
The work of M.H. is supported in part by the Grant-in-Aid of the Japanese Ministry 
of Education, Sciences and Technology, Sports and Culture (MEXT) 
for Scientific Research (No. 25287046). 
The work of N.I. is supported in part by JSPS KAKENHI Grant Number 25800143. 
The work of M.H. and N.I. were also supported in part by the National Science Foundation under Grant No. PHYS-1066293 and the hospitality of the Aspen Center for Physics.
\appendix


\end{document}